\documentclass[twocolumn,prd,nofootinbib,superscriptaddress,longbibliography]{revtex4} %

\newcommand\ForInternalReference[1]{}
\newcommand\SkipForEarlyCirculation[1]{}

\newcommand\SkipPP[1]{}

\usepackage[colorlinks=true,citecolor=blue,urlcolor=blue]{hyperref}
\usepackage{verbatim}
\usepackage{graphicx}
\usepackage{dcolumn}
\usepackage{bm}
\usepackage{color}
\usepackage{xcolor}
\usepackage{xspace}
\usepackage{url}
\usepackage{mathtools}
\usepackage{float}
\usepackage{multirow}
\usepackage{adjustbox}
\usepackage[normalem]{ulem}
\usepackage{bm}
\usepackage{placeins}
\usepackage{afterpage}
\usepackage{amsmath,amssymb}
\usepackage{acronym}
\newcommand\optional[1]{}
\acrodef{NR}[NR]{Numerical Relativity}

\newcommand{\NRSur}{NRHybSur3dq8\xspace}

\usepackage{color}
\definecolor{amber}{rgb}{1.0, 0.75, 0.0}
\definecolor{orange}{rgb}{1.0, 0.5, 0.0}
\definecolor{amaranth}{rgb}{0.9, 0.17, 0.31}

\graphicspath{{./Figures/}}
\newcommand{\mc}{{\cal M}}

\def\ltsima{$\; \buildrel < \over \sim \;$}
\def\simlt{\lower.5ex\hbox{\ltsima}}
\def\gtsima{$\; \buildrel > \over \sim \;$}
\def\simgt{\lower.5ex\hbox{\gtsima}}

\newcommand{\UT}{\affiliation{Center for Gravitational Physics, The University of Texas at Austin, Austin, Texas 78712, USA}}

\begin{document}

\title{Accuracy limitations of existing numerical relativity waveforms on the data analysis of current and future ground-based detectors}

\author{Aasim Jan}
\UT
\author{Deborah Ferguson}
\UT
\affiliation{Illinois Center for Advanced Studies of the Universe, The University of Illinois Urbana-Champaign, Urbana, Illinois 61801, USA}
\author{Jacob Lange}
\UT
\author{Deirdre Shoemaker}
\author{Aaron Zimmerman}
\UT

\begin{abstract}

As gravitational wave detectors improve in sensitivity, signal-to-noise ratios of compact binary coalescences will dramatically increase, reaching values in the hundreds and potentially thousands. Such strong signals offer both exciting scientific opportunities and pose formidable challenges to the template waveforms used for interpretation. Current waveform models are informed by calibrating or fitting to numerical relativity waveforms and such strong signals may unveil computational errors in generating these waveforms. In this paper, we isolate a single source of computational error, that of the finite grid resolution, and investigate its impact on parameter estimation for aLIGO and Cosmic Explorer. We demonstrate that increasing the inclination angle or decreasing the mass ratio $q$ ($q \leq 1 $) raises the resolution required for unbiased parameter estimation. We quantify the error associated with the highest-resolution waveform utilized in our study using an extrapolation procedure on the median of recovered posteriors and confirm the accuracy of current waveforms for the synthetic sources. We introduce a measure to predict the necessary numerical resolution for unbiased parameter estimation and use it to predict that current waveforms are suitable for equal and moderately unequal mass binaries for both detectors. However, current waveforms fail to meet accuracy requirements for high signal-to-noise ratio signals from highly unequal mass ratio binaries $(q \lesssim 1/6)$, for all inclinations in Cosmic Explorer, and for high inclinations in future updates to LIGO. Given that the resolution requirement becomes more stringent with more unequal mass ratios, current waveforms may lack the necessary accuracy, even at median signal-to-noise ratios for future detectors.

\end{abstract}
\maketitle

\section{Introduction}
\label{sec:Intro}
The detection of gravitational waves (GW) \cite{PhysRevX.9.031040,PhysRevX.11.021053,2021arXiv210801045T,2021arXiv211103606T} by the Advanced Laser Interferometer Gravitational-wave Observatory (LIGO) \cite{2015CQGra..32g4001L} and Advanced Virgo \cite{2015CQGra..32b4001A} has resulted in an explosive growth in the field of GW astronomy and this growth is expected to continue due to plans for substantial upgrades to current detectors \cite{Abbott_2020,Insrument_science_paper}  and the development of next-generation ground-based \cite{Punturo_2010,Abbott_2017,reitze2019program,reitze2019cosmic,evans2021horizon} and space-based \cite{Luo_2016,amaroseoane2017laser,babak2021lisa} detectors over the next two decades. These advancements will enable the detection of GW events at significantly higher signal-to-noise ratios (SNRs) than observed to date. 
Detection of such strong signals not only promises unprecedented science but also raises a question: Are current waveforms accurate enough to yield unbiased parameter estimation (PE) of GW sources? Increased sensitivities necessitate a stricter accuracy requirement on model waveforms and the numerical relativity (NR) waveforms used in their construction. The increase in the accuracy requirement stems from the fact that as the SNR of observed signals increases, the statistical uncertainty in the recovered parameters decreases, amplifying the impact of underlying systematic uncertainties. As such, to fully extract the wealth of information contained in such strong signals, it is imperative that the waveforms meet these enhanced accuracy requirements.

A coalescing compact binary system is described by Einstein's field equations, which can be solved using NR \cite{Pretorius_2005,Baker_2006,Campanelli_2006}, particularly for the late inspiral, merger, and ringdown.
Despite being the most accurate way of obtaining a gravitational waveform of merger, NR waveforms are typically not used directly for GW data analysis as they are computationally expensive to generate, are short in length, and have sparse and uneven parameter space coverage. 
While some studies have used NR waveforms directly for PE \cite{Lange_2017,2021PhRvL.126h1101B} including a LIGO-Virgo-KAGRA analysis of GW150914 \cite{PhysRevD.94.064035}, these analyses are limited by the number of available simulations. 
As such, to create comprehensive inspiral-merger-ringdown waveform models that cover a continuous parameter space, various modeling strategies have been developed. 
These include: effective-one-body formalism \cite{PhysRevD.59.084006,PhysRevD.62.064015,PhysRevD.62.084011,PhysRevD.64.124013},  phenomenological modeling \cite{Ajith_2007,PhysRevD.77.104017,PhysRevLett.106.241101}, and NR surrogates \cite{PhysRevLett.115.121102}. Each of these modeling strategies relies on NR waveforms. 
Effective-one-body models
use NR waveforms to calibrate the free parameters, phenomenological models 
use hybridized NR and post-Newtonian waveforms \cite{blanchet_gravitational_2006} for a multi-parameter fit and NR surrogates 
generate interpolants based on NR waveforms. These waveform models play a critical role in the search and interpretation of GW events, and given that their accuracy is bounded by the accuracy of NR waveforms, it is imperative that the NR waveforms used in their construction meet the required level of accuracy. 

Both NR and model waveforms have inherent uncertainties that can affect the interpretation of GW signals. 
One of the sources of systematic uncertainty in NR waveforms is inaccuracy from finite grid resolution.
The uncertainties in model waveforms arise from the approximations made during their construction, omission of certain physics and their reliance on finite-resolution NR waveforms. 
However, it is worth noting that it has been shown that currently available NR waveforms demonstrate an adequate level of accuracy for the signals detected to date \cite{PhysRevD.104.044037}.

While multiple studies \cite{PhysRevD.96.124041,PhysRevResearch.2.023151,PhysRevD.102.124069,2022PhRvD.106d4042H,owen2023waveform} have been done to investigate the impact of model waveform systematics on PE, an area that remains relatively unexplored is the effect of bias introduced by finite-resolution NR waveforms on full PE. In a previous study \cite{Lange_2017}, one-dimensional PE of an NR injection was conducted to study the impact of numerical resolutions on the final posterior distribution, with the recovery being done using NR waveforms. Apart from being restricted to a single dimension, the study was carried out at an SNR of $25$, 
limiting the applicability of such a study to future, more sensitive detectors. In another study \cite{PhysRevD.104.044037}, a criterion was proposed that relates the minimum resolution required for producing an NR waveform that is indistinguishable \cite{PhysRevD.82.084020} from a true signal to the SNR. 
This criterion is conservative, as indistinguishability itself is an unnecessarily strict requirement \cite{Chatziioannou:2017tdw,PhysRevResearch.2.023151} and significant parameter bias may not arise even for technically distinguishable waveforms. Consequently, its application can become impractical, especially in scenarios where higher modes play a crucial role as the criterion sets a considerably higher resolution requirement than actually needed, placing a burden on computational resources.

In this work, we study the impact of using finite-resolution NR waveforms on PE by performing PE on synthetic signals (injections) produced by NR simulations run at multiple resolutions. We start by comparing the PE outputs of equal mass ratio binary injections differing only in resolution, and investigate the impact of NR truncation errors on the recovered binary parameters. We demonstrate that using a resolution higher than what is required for unbiased PE does not yield additional information for PE. We confirm that the resolution requirement based on waveform indistinguishability \cite{PhysRevD.78.124020} is stricter than required for unbiased PE. We show the accuracy requirement for NR waveforms is dependent on the detector, as the impact of truncation errors varies according to the sensitivity curve of the detector. We show that for systems with a greater higher mode content in their GWs, higher resolution is needed to achieve unbiased PE than for those without.
Moreover, we introduce and employ an extrapolation procedure to estimate errors associated with the highest resolution waveforms used in our work. Finally, we utilize our results, combined with a waveform criterion, to make predictions for future NR codes, predicting the minimum resolution necessary for unbiased PE of signals expected to be observed by upcoming detectors. It should be noted that the results of this study apply to NR codes that use finite differencing methods to solve Einstein's equations.

The rest of the paper is organized as follows: in Sec.~\ref{sec:methods} we review the NR waveforms used in the study, the indistinguishability criterion for NR waveforms generated using finite-differencing codes, and the PE code {\tt RIFT}. 
In Sec.~\ref{sec:PEresults} we study the impact of numerical biases on PE across a variety of systems and for two detectors, LIGO Hanford (H1) and the proposed third generation detector Cosmic Explorer (CE) \cite{reitze2019cosmic}. 
In Sec.~\ref{sec:extrapolation}, we employ an extrapolation procedure to determine bias in PE due to the highest resolution waveforms used in this study. 
In Sec.~\ref{sec:prediction}, we predict the SNR at which we will see significant biases and compare them with what we obtained from full PE. 
We also assess the accuracy of current waveforms and determine if they are sufficiently accurate for current and future detectors. 
In Sec.~\ref{sec:conclude} we summarize our results.

\section{Methods}
\label{sec:methods}

In this section, we introduce the methods we use to investigate the effect of NR truncation errors on PE and assess the accuracy limitations of existing NR waveforms in the context of PE. 
We start by discussing the NR waveforms used as injections in our PE studies in Sec.~\ref{ssec:NR}. 
We then discuss the indistinguishability criterion for NR waveforms, as proposed in \cite{PhysRevD.104.044037}, 
in Sec.~\ref{ssec:criteria}. 
We then review the PE algorithm {\tt RIFT} in section \ref{ssec:RIFT}. 

\subsection{Numerical Relativity waveforms} 
\label{ssec:NR}

\begin{table*}
    \centering
    \addtolength{\tabcolsep}{5pt} 
     \begin{tabular}{c c c c c c c c}
     \hline
         $q$ & $m_1/M_\odot$ & $m_2/M_\odot$ & $\mathcal{M}_c$/$M_\odot$ & $\chi_{1}$ &$\chi_{2}$& $\iota$ (radians) & $\Delta$ \\ [0.5ex] 
         \hline\hline
         
         \vspace{0.25cm}
         
         1 & 50 & 50  & 43.5  & (0.0, 0.0, 0.6) & (0.0, 0.0, 0.6)& $0$ & $M/80$, $M/120$, $M/140$, $M/200$  \\ 

         \vspace{0.25cm}     
         
         1 & 50 & 50  & 43.5  & (0.0, 0.0, 0.6) & (0.0, 0.0, 0.6)& $\pi/6$ & $M/80$, $M/120$, $M/140$, $M/200$  \\

         1/3 & 112.5  & 37.5 & 54.9  & (0.0, 0.0, 0.0) &(0.0, 0.0,  0.0) & $0$ &  $M/100$,  $M/120$, $M/140$, $M/180$ \\[1ex] 
     \hline
     \end{tabular}
     \addtolength{\tabcolsep}{-13pt} 
     \caption{\textbf{Synthetic sources}: Parameters of the three sets of synthetic sources used in our study. 
     For these sources, we have chosen right ascension ra = $0.57$, polarization angle $\psi = 0$, declination dec = $0.1$, and coalescence phase $\phi_c = 0$ (all in radians). The GPS time at the geocenter was set to $10^{9}$s. 
     Each set of injections was recovered at multiple SNRs, achieved by changing the luminosity distance $D_L$.}
     \label{tab:param}
\end{table*}

For our PE investigations, we use NR waveforms generated using the {\tt MAYA} code \cite{Herrmann_2007,PhysRevD.76.084020,PhysRevLett.103.131101,PhysRevD.88.024040} as injections.  
{\tt MAYA} is a branch of the {\tt Einstein Toolkit}, and is built upon the {\tt Cactus} framework, incorporating {\tt Carpet} \cite{Schnetter_2004} for mesh refinement. It employs the BSSN \cite{PhysRevD.59.024007} formulation to derive the initial constraints and evolution equations from Einstein's field equations. In its calculations, {\tt MAYA} utilizes sixth-order spatial finite-differencing and fourth-order Runge-Kutta for time evolution.

Our NR waveform injections were extracted from two sets of quasi-circular {\tt MAYA} NR simulations. 
Each had identical initial conditions and parameters, differing only in grid resolution.
These simulations are parametrized, up to an arbitrary total mass $M$ (in code units), by the intrinsic parameters ${\bm \lambda}$ of the binary.
These intrinsic parameters include primary mass $m_1$, secondary mass $m_2$, and spin vectors $\bm{S}_1$ and $\bm{S}_2$.
Since the simulations are scale-invariant with respect to total mass, it is conventional to parametrize them in terms of their mass ratio $q = m_2/m_1 \leq 1$.
Further, we define the dimensionless spin parameters ${\bm \chi}_1 = {\bm S}_1/m_1^2$ and ${\bm \chi}_2 = {\bm S}_2/m_2^2$. 

The first set of simulations had $q=1$ and spins aligned with the orbital angular momentum, with $\chi_{1z} = \chi_{2z} = 0.6$.
The other set was non-spinning, with $q = 1/3$.
These parameters were computed at the beginning of the simulation, but there is evidence that non-physical junk radiation does not significantly affect their values \cite{PhysRevD.100.081501}. 

The initial separations and grid spacing or resolution of our simulations are expressed in terms of $M$.
For the $q=1$ systems, the initial separation was $12M$ while for the $q=1/3$ systems, the initial separation was $9M$. 
Both sets consist of four differently resolved simulations. 
The simulations 
were performed on a grid with $10$ refinement levels with the largest grid radii being $409.6 M$ and the smallest being $0.2 M$. 
The resolutions are specified by the grid spacing $\Delta_i$ of the finest grid in each simulation and varied from $\Delta_i = M/200 ~(M/180)$ for our highest resolution $q=1~(q=1/3)$ simulation to $\Delta_i = M/80~(M/100)$ for our lowest.
The lowest resolution in each set is lower than typically used for GW data analysis and was chosen to illustrate the impact of resolution at moderate SNRs.
The typical resolution of $q=1$ and $q=1/3$ simulations in the {\tt MAYA} catalog~\cite{Ferguson:2023vta} is $M/200$ and $M/370$ respectively.

The waveforms were computed from the Weyl scalar $\Psi_4$ extracted at a finite radius of $75M$  from the binary system. We avoided extrapolating the waveforms to infinity in order to isolate the impact of finite resolution. 
$\Psi_4$ is related to the gravitational wave polarizations as follows:
\begin{equation}
    \Psi_4(t) = \ddot{h}_+(t) - i\ddot{h}_\times(t) \,.
\end{equation}
Further decomposition of GW polarizations involves expressing them as a sum of spherical harmonics $_{-2}Y_{lm}$ and GW modes $h_{lm}$, given by
\begin{equation}
    h_+(t) - ih_{\times}(t) = \sum_{l,m} h_{lm}(t)~_{-2}Y_{lm}\,.
\end{equation}
In our injections, we only used $\ell\leq4$ modes.

\begin{figure*}
    \includegraphics[scale=0.45]{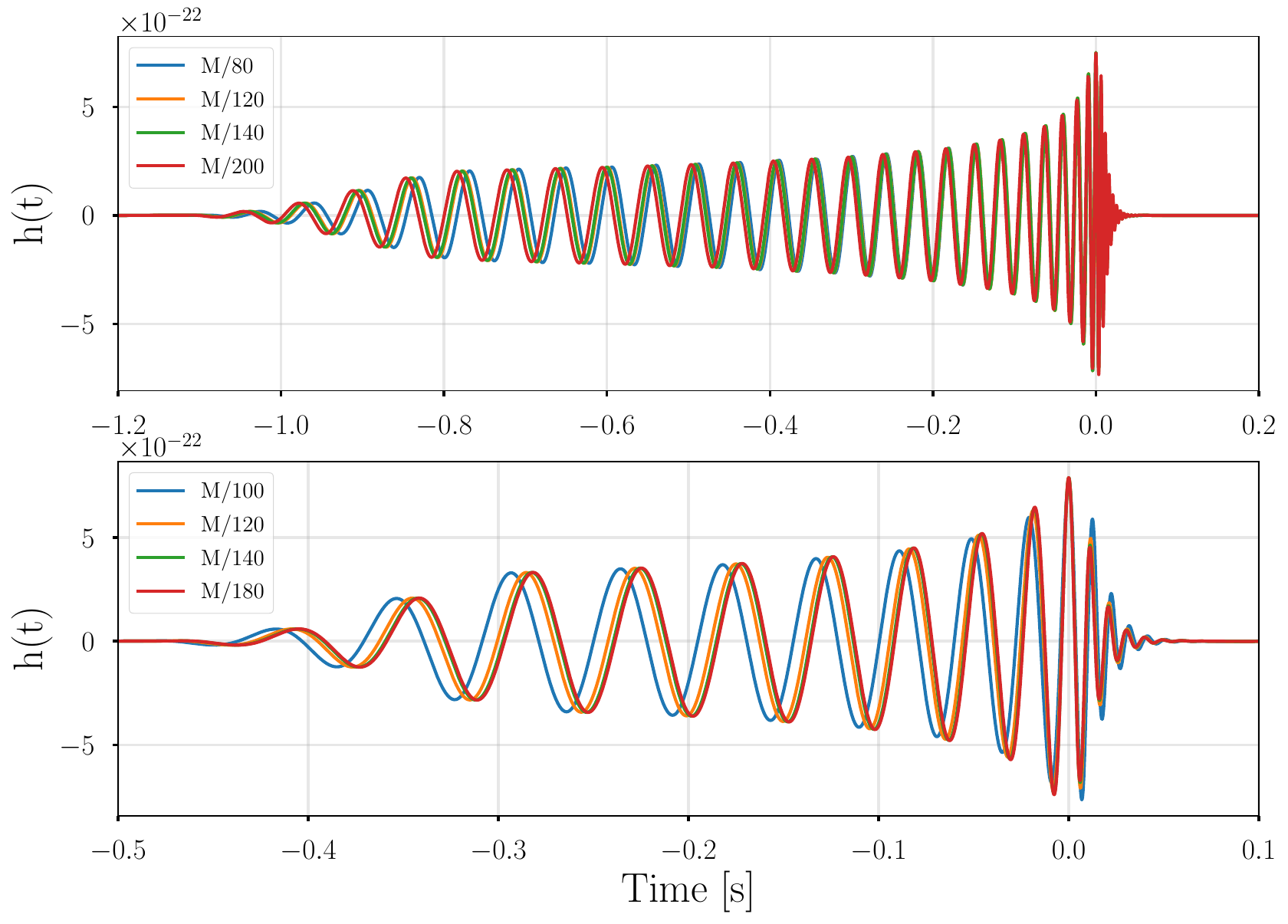}
    \caption{\textbf{Comparison between the different resolutions of $q=1$ and $q=1/3$ waveforms}: \textit{Top panel}: 
    Strain evaluated from the four distinct resolution $q=1$ waveforms for a source with $M_{\text{tot}} = 100 M_\odot, \iota = 0$. \textit{Bottom panel}: Strain evaluated from the four different resolution $q=1/3$ waveforms for a source with $M_{\text{tot}} =  150 M_\odot, \iota = 0$. In both panels, the waveforms are aligned at the peak, evaluated at  $D_L = 1000$ Mpc and the colors represent different resolutions.}
    \label{fig:waveforms}
\end{figure*}

In order to compute the GW strain $h(t)$ as measured at a detector, we must specify the detector frame total mass $M_\text{tot}$ so that the component masses and dimensionful scales are determined.
Additionally, the extrinsic parameters of the binary must be defined. 
These extrinsic parameters determine the space-time location and orientation relative to the detector and include luminosity distance $D_L$, right ascension ra, declination dec, polarization $\psi$, inclination $\iota$, orbital phase $\phi_c$, and coalescence time $t_c$.
The intrinsic and extrinsic parameters chosen for our NR waveforms, along with the grid spacing used in the finest grid, are provided in Table~\ref{tab:param}.
There and throughout this study we quote the detector-frame masses for the system.
Two sets of these waveforms are assumed to be observed face-on, with $\iota=0$, while the third has a modest inclination $\iota = \pi/6$.
Regardless of the detector sensitivity used in this study, we assume the detector is located at the site of the Hanford GW detector and aligned with it.
Figure~\ref{fig:waveforms} depicts the time-domain strains imprinted on the detector for the face-on cases.
These waveforms are aligned at their peaks, and we can see a clear dephasing with differing grid resolutions.

\subsection{Criterion for assessing waveform accuracy}
\label{ssec:criteria}
A gravitational waveform $h_i$ extracted from an NR simulation can be expressed in terms of the exact solution $h$ as $h_i = h +\delta h_i$, where $\delta h_i$ is the error in the extracted waveform. 
If the code used to generate the waveform uses finite-differencing to solve partial differential equations, then the truncation error can be expressed as $\delta h_i = c(\Delta_i)^\alpha$, where $\alpha$ is the convergence rate of the code and $c$ depends on the derivatives of $h$.
Recall that since {\tt Carpet} uses adaptive mesh refinement, different sections of the grid have different spacings, and $\Delta_i$ refers to the spacing of the finest grid.

Given two waveforms $h_1$ and $h_2$, the overlap between them is defined as
\begin{equation}
    \mathcal{O}[h_1,h_2] = \frac{\langle h_1|h_2 \rangle} {\sqrt{\langle  h_1|h_1 \rangle \langle  h_2|h_2\rangle}} \,,
\end{equation}
where the noise-weighted inner product is defined as
\begin{equation}
    \langle   h_1|h_2 \rangle = 4 \operatorname{Re} \int_{f_{\rm min}}^{f_{\rm max}}  \frac{h_1^*(f) h_2(f)}{S_n(f)}df \,, 
    \label{eq:IP_Mod}
\end{equation}
with $S_n(f)$ being the one-sided power spectral density of the detector, $f_{\rm min}$ a low-frequency cutoff, $f_{\rm max}$ a high-frequency cutoff, and $*$ denoting the complex conjugate. 
Here, we have used the fact that $h_i(t)$ is a real-time series, and as such its Fourier transform satisfies $h_i(f) = h^*_i(-f)$, allowing us to define the inner product as an integral over positive frequencies.

Using the overlap, we can define mismatch $\epsilon[h_1,h_2]$, which several previous investigations have argued relates to systematic biases in PE
\cite{PhysRevD.78.124020,PhysRevD.79.124033,PhysRevD.82.084020,PhysRevD.87.024004,PhysRevD.82.124052,PhysRevD.93.104050,PhysRevResearch.2.023151}, as
\begin{align}
\epsilon[h_1,h_2] = 1 - \max_{t_c,\phi_c} \mathcal{O}[h_1,h_2] \,.
\end{align}
We calculate mismatch between two NR waveforms, both having identical parameters but differing in the numerical resolution of simulation grid. We use this mismatch to compute $\beta$ which is a parameter that only depends on the parameters of the binary system and can be computed as
\begin{equation}
    \epsilon[h_1,h_2] = \frac{\beta^2}{2} (\Delta_2^\alpha - \Delta_1^\alpha )^2 \,.
    \label{eq:beta}
\end{equation}
Using $\beta$, and the convergence rate $\alpha$ of the finite-differencing code, we can then estimate the minimum resolution necessary for producing NR waveforms indistinguishable from a true signal using the criterion \cite{PhysRevD.104.044037}
\begin{equation}
    \Delta < (\rho \beta)^{-1/\alpha} \,,
    \label{eqn:criteria}
\end{equation}
where $\rho$ is the SNR.
Ideally, $\alpha$ and $\beta$ would be independent of the resolution of the simulations used for their calculation and this would be true if the waveforms were extracted from simulations that consisted of a single grid, used the same order for all finite-differencing, and the strain was a grid variable. 
However, given the complicated mesh refinements and boundary conditions as well as the interpolation necessary to extract $\Psi_4$ and compute strain, we have a less well-defined convergence order. 
This means the values of $\alpha$ and $\beta$ differ slightly between different pairs of resolutions. 
This introduces an uncertainty of up to $10 M^{-1}$ in our estimates of the minimum resolution necessary for indistinguishability.

\subsection{RIFT}
\label{ssec:RIFT}
A GW from a quasicircular binary black hole system undergoing merger can be completely determined by 15 parameters. 
These parameters are classified into the two groups, \textit{intrinsic} ($\bm{\lambda}$) and \textit{extrinsic} ($\bm{\theta}$), described previously. 
In discussing the PE results, we also use chirp mass $\mc_c = (m_1 m_2)^{3/5}/(m_1+m_2)^{1/5}$, which is paired with the mass ratio $q$ to represent the mass parameters. 
We also define another parameter called effective spin, which  is a mass-weighted combination of individual spins and is defined as
\begin{equation}
    \chi_{\rm eff} = \frac{m_1 \chi_{1z} + m_2 \chi_{2z}}{m_1+m_2} \,.
\end{equation}

After a GW is detected, the data $d$ is analyzed to infer the parameters of the radiating system using a PE algorithm, and {\tt RIFT} \cite{gwastro-PENR-RIFT} is one such algorithm. It is a highly parallelizable, grid-based, iterative algorithm consisting of two core iterative stages. 
In the first stage, a ``grid" of intrinsic parameters is put forth, and for each point $\bm{\lambda}_\alpha$ from the proposed grid, {\tt RIFT} integrates over the extrinsic variables to compute the marginal likelihood 
\begin{equation}
     {\cal L}_{\rm marg}(d| {\bm \lambda}) \equiv 
     \int {\cal L}_{\rm full}(d|\bm{\lambda},{\bm \theta})
     p({\bm \theta})d{\bm \theta}
\end{equation}
from the likelihood ${\cal L_{\text{full}}}(d|\bm{\lambda} ,{\bm \theta} )$ of the GW signal,
accounting for detector response. 
Here, $p({\bm \theta})$ are the priors on the extrinsic parameters.
The integration is made possible by factorizing the dependence of $\mathcal L_{\rm full}$ on the extrinsic parameters, 
which is partially made possible by expressing the GW polarizations in terms of GW $\ell, m$ modes; see~\cite{Pankow:2015cra} for a more detailed specification.  

Once the marginalized likelihood is evaluated for points on the grid $(\bm{\lambda}_\alpha,{\cal L}_\alpha)$, {\tt RIFT} interpolates this discrete grid of marginalized likelihood points to generate the continuous likelihood distribution ${\cal L}_{\rm marg}(d|\bm{\lambda})$. 
With the knowledge of the continuous marginalized likelihood distribution ${\mathcal L}_{\rm marg}(d|\bm{\lambda})$ and intrinsic prior $p(\bm{\lambda})$, {\tt RIFT} constructs the marginalized posterior via Bayes' Theorem:
\begin{equation}
\label{eq:post}
    p_{\rm post}({\bm \lambda}|d)=\frac{{\cal L}_{\rm marg}(d|\bm{\lambda} )p(\bm{\lambda})}{\int d\bm{\lambda} {\cal L}_{\rm marg}(d|\bm{\lambda} ) p(\bm{\lambda} )}.
\end{equation}
The integral in the denominator is calculated by
performing a Monte Carlo integral:  the evaluation points and weights in that integral are weighted posterior samples,
which are fairly resampled to generate conventional independent, identically distributed posterior samples.

The grid for the following iteration is generated using a subset of posterior samples from the previous iteration, with an additional expansion of the grid to ensure that regions of high likelihood that might have been missed can be explored.
The iterations continue until two successive iterations converge, as determined by examining the Jensen-Shannon (JS) divergence \cite{JS_test} between the one-dimensional marginal posteriors for each intrinsic parameter.
In our study, we have taken measures to guarantee convergence by ensuring that the JS divergence between the last two iterations is less than or equal to $10^{-3}$ for all parameters.
Further details and a justification of this choice is given in Appendix~\ref{sec:JS}.
For further details on {\tt RIFT}'s technical underpinnings, performance, and comparison to other PE codes see \cite{Pankow:2015cra,gwastro-PENR-RIFT,gwastro-PENR-RIFT-GPU,gwastro-mergers-nr-LangePhD, Wofford:2022ykb}.

\section{Parameter estimation results}
\label{sec:PEresults}

In this section, we present the results of our PE study.
Our goal is to investigate the impact of finite-resolution errors in NR waveforms on PE recovery.
Ideally, to achieve this goal, we would inject an infinite-resolution NR waveform followed by recovery using identically resolved NR waveforms as template waveforms. 
This process would be iterated multiple times, with each iteration utilizing template waveforms obtained from NR simulations of differing numerical resolution.
Ultimately, we would compare the results of each iteration and assess the impact of truncation errors on the posterior distributions across a range of SNRs.
While the grid-based structure of {\tt RIFT} allows it to perform PE using NR waveforms \cite{Lange_2017}, the available catalog of NR waveforms are all of varying resolution, preventing us from isolating the effect of finite-resolution. 
As such we must take a different approach, which we describe in Sec.~\ref{ssec:Approach}.
We then describe the setup of our PE study in Sec.~\ref{ssec:Setup}, and present results for our three sets of NR injections (Tab.~\ref{tab:param}) in Secs.~\ref{ssec:q1}, \ref{ssec:q1inc}, and \ref{ssec:q3}.

\subsection{Justification for PE approach}
\label{ssec:Approach}

Given the impracticality of utilizing NR waveforms as our model waveform for PE, we approach the problem differently.
We consider a sequence $h_i$ of NR simulations with varying resolution but fixed binary parameters ${\bm \mu}_*$, which we inject into a zero-noise realization for use as our synthetic data $d_i$.
Next we use a high-accuracy waveform model, \NRSur ($h_{\rm sur}({\bm \mu})$) for parameter recovery, one which can be evaluated at arbitrary parameters ${\bm \mu}$.
The result is a sequence of posteriors $p({\bm \mu} | d_i, h_{\rm sur})$ under our model hypothesis.
We repeat this procedure with a variety of SNR values for our injected waveform, and compare the resulting posteriors across resolution values in order to understand the effect of finite resolution on PE results.

\begin{figure}[tb]
\includegraphics[width=0.98\columnwidth]{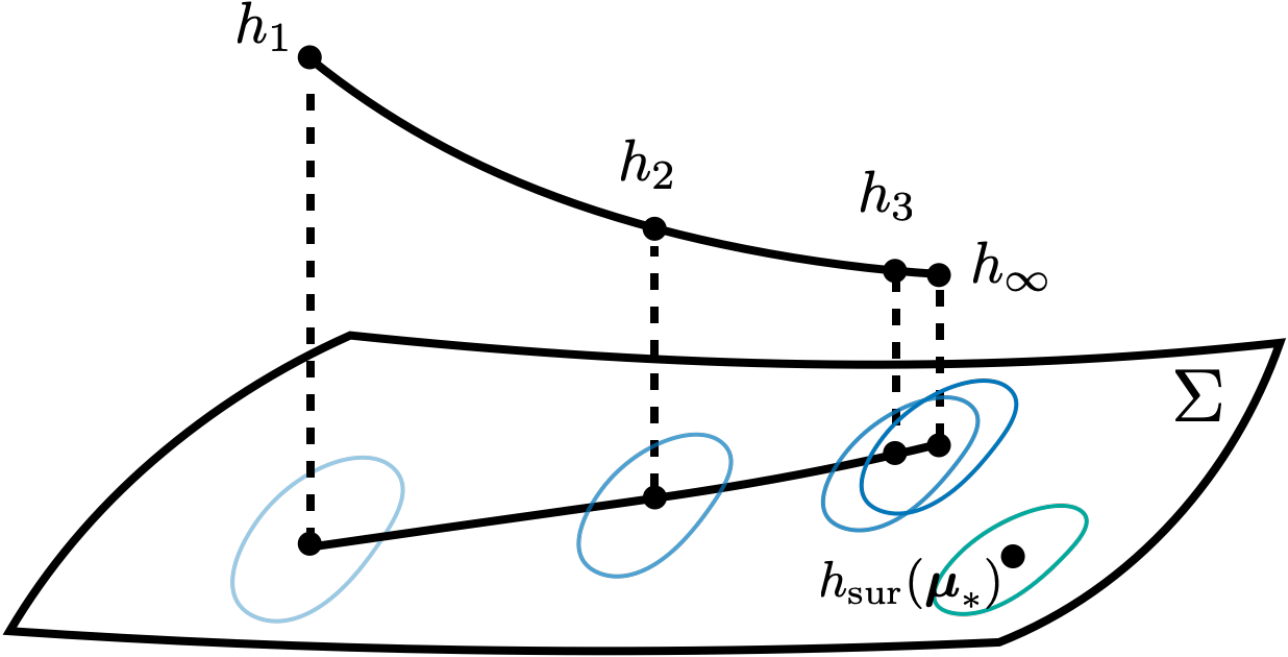} 
\caption{Illustration of our data analysis strategy, visualizing the waveforms as points in signal space. 
We perform PE on a sequence of NR waveforms $h_i$ at fixed binary parameters ${\bm \mu}_*$ and recover with a waveform model $h_{\rm sur}$, which produces waveforms on a submanifold $\Sigma$. 
As we increase the resolution, the waveforms converge to an extrapolated signal $h_\infty$, and the sequence of posteriors also converges; the manner in which these posteriors converge as a function of injected SNRs allows us to determine resolution requirements for unbiased PE. Meanwhile, $h_{\rm sur}({\bm \mu}_*)$ is a separate signal which lies off this sequence in general.}
\label{fig:Resolution}
\end{figure}

Figure~\ref{fig:Resolution} illustrates how we can use these results to explore this question when we recover our posteriors with the model $h_{\rm sur}$.
The figure illustrates the space of possible signals.
All signals that can be realized by the model $h_{\rm sur}$ when evaluated over the relevant range of parameter values ${\bm \mu}$ forms a submanifold $\Sigma$ of the overall signal space. 
The numerical simulations $h_i$ do not lie on $\Sigma$, but using the match as a distance measure allows us to identify points ${\bm \mu}_i$ such that $h_{\rm sur}({\bm \mu}_i)$ is the best match to $h_i$.
The distance of each $h_i$ away from $h_{\rm sur}({\bm \mu}_i)$ is a measure of the SNR loss due to mismodeling the NR waveform with $h_{\rm sur}$.
The difference between the best fit ${\bm \mu}_i$ value and ${\bm \mu}_*$ is the parameter bias that we are interested in.
The sequence of NR waveforms converges to some extrapolated waveform $h_\infty$, which differs from $h_{\rm sur}({\bm \mu}_*)$.

From the perspective of PE, the best fit points ${\bm \mu}_i$ provide the maximum likelihood values for a zero-noise recovery with the waveform model $h_{\rm sur}$.
$h_{\rm sur}({\bm \mu}_i)$ provides the relevant maximum likelihood waveform.
Meanwhile, the shape of the posteriors is determined primarily as if $h_{\rm sur}({\bm \mu}_i)$ were the injected data, modulo the effects of SNR loss due to the projection of $h_i$ onto $\Sigma$.
Thus we can study the sequence of recovered posteriors as we approach $h_\infty$ and the corresponding recovered posteriors, and ask at what SNR and for which pairs of resolutions the PE results are indistinguishable.
This allows us to state what approximate resolution is required at a given SNR for unbiased parameter recovery.
Potentially the most important complicating factor relative to an idealized study is the variable SNR loss as we move along the sequence $h_i$; if this SNR loss is not too severe, this approach provides the desired information about the PE bias from finite resolution.

\subsection{Setup of PE study}
\label{ssec:Setup}

\begin{figure}[t]
    \includegraphics[width = \columnwidth]{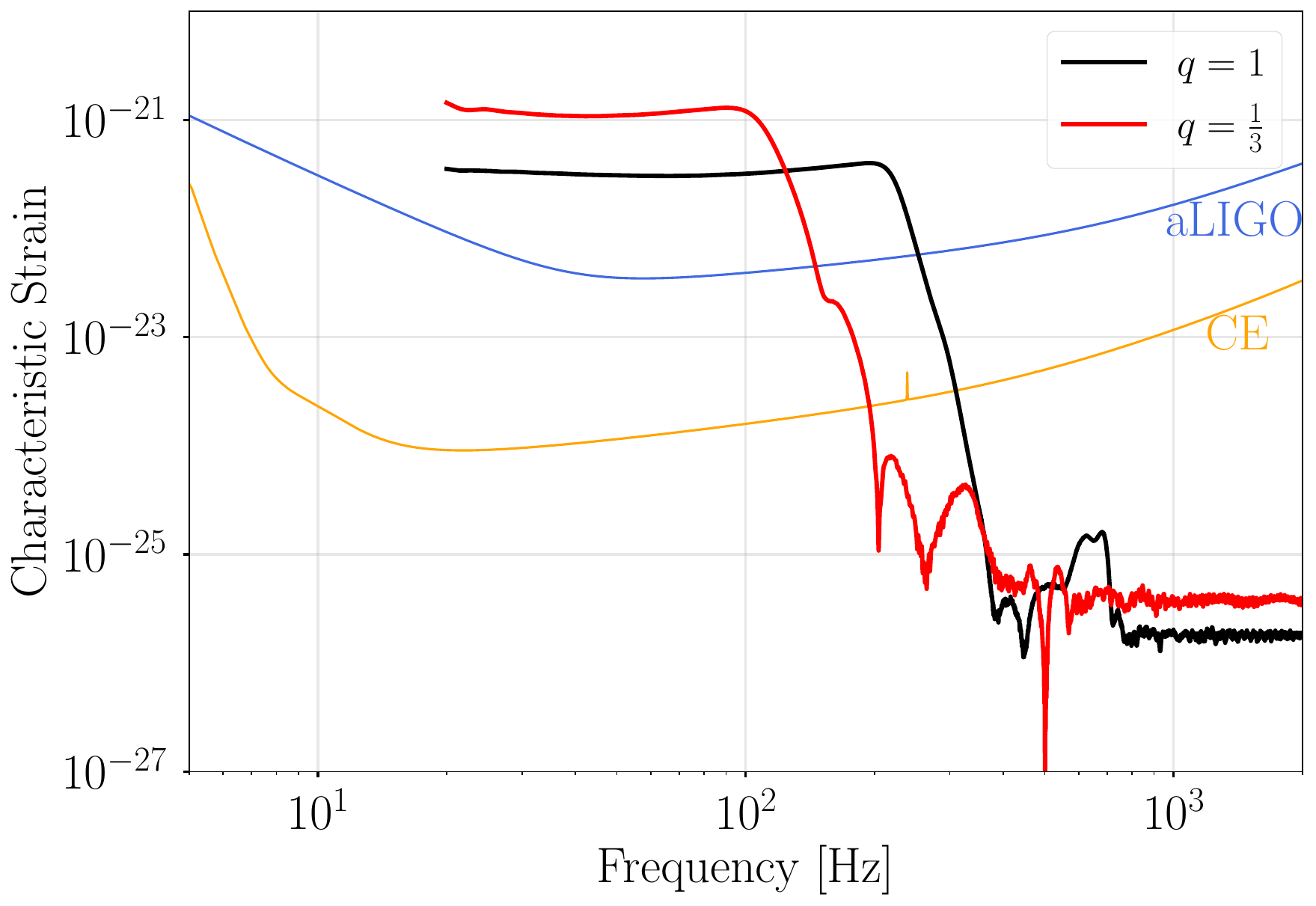}
    \caption{\textbf{Characteristic strains and noise amplitudes}: Characteristic strain evaluated from the highest resolution $q=1$ (black) and $q=1/3$ waveform (red). Black curve is from a source with parameters $M_{\text{tot}} =  100 M_\odot, \iota = 0$, SNR $=12$, and red curve is from a source with parameters $M_{\text{tot}} = 150 M_\odot, \iota = 0,$ SNR $=37$. 
    Also plotted are the noise amplitudes for design aLIGO and CE. The extrinsic parameters for both waveforms are as given in Table~\ref{tab:param}.}
    \label{fig:ch_strain}
\end{figure}

To investigate the impact of finite-resolution on PE, we first injected four different resolutions of $q = 1$ gravitational waveforms (refer to Table \ref{tab:param} for all injected parameters) at a range of SNRs. 
Instead of directly selecting SNR values, we first selected eight different resolutions and then found the corresponding critical SNR using Eq.~\eqref{eqn:criteria}. Going forward, when referring to
critical SNR and resolution, we mean the SNR and resolution values at which
the inequality transitions to equality. The choice for resolution levels was made strategically such that for the first four SNR values, two would correspond to cases where $M/80$ waveform is predicted to be indistinguishable from true signal and two to cases where it is predicted to be distinguishable from true signal. 
We applied the same approach for $M/120$ to select the remaining four SNR values. 
The selected resolutions and corresponding critical SNRs are given in the first two columns of Table~\ref{tab:q1_H1_bias}.
Subsequently, we analyzed and compared the resulting posterior distributions after recovery with the hybridized NR surrogate waveform model \NRSur \cite{PhysRevD.99.064045} using all available $\ell\leq4$ modes, and the {\tt RIFT} PE code, quantifying the impact of truncation errors as a function of SNR.  

\begin{table*}[tb]
\centering
\addtolength{\tabcolsep}{20pt} 
     \begin{tabular}{c c c c c}
     \hline \
     SNR & $\Delta_\text{critical}$ & $M/80$ & $M/120$ & $M/140$\\ [0.5ex] 
     \hline\hline
     9 & $M/70$ & -0.28 & -0.04 & -0.02  \\ 
     12 & $M/75$ & -0.39 & -0.07 & -0.03 \\
     20 & $M/85$ & -0.66 & -0.12 & -0.06 \\
     31 & $M/95$ & -1.09 & -0.17 & -0.09 \\
     47 & $M/105$  & -1.70 & -0.27 & -0.14 \\
     67 & $M/115$  & -3.10 & -0.49 & -0.22 \\
     94 & $M/125$ & -7.89 & -0.92 & -0.25 \\
     128 & $M/135$ & -9.59 & -1.22 & -0.78 \\[1ex] 
     \hline \\

     \hline 
     SNR & $\Delta_\text{critical}$ & $M/80$ & $M/120$ & $M/140$\\ [0.5ex] 
     \hline\hline
     5 & $M/70$ & -0.10 & -0.02 & -0.01  \\ 
     7 & $M/75$ & -0.28 & -0.05 & -0.03 \\
     11 & $M/85$ & -0.55 & -0.11 & -0.05 \\
     16 & $M/95$ & -0.89 & -0.17 & -0.08 \\
     22 & $M/105$  & -1.21 & -0.24 & -0.09 \\
     31 & $M/115$  & -1.74 & -0.32 & -0.11 \\
     44 & $M/125$ & -2.53 & -0.52 & -0.22 \\
     60 & $M/135$ & -3.58 & -0.69 & -0.28 \\[1ex] 
     \hline

     \end{tabular}
     \addtolength{\tabcolsep}{-20pt}
     \caption{\textbf{Normalized bias in the marginalized $\mc_c$ posterior distributions of $\boldsymbol{q=1, \iota = 0}$ injections}: Bias observed in the lower resolution posteriors, calculated with respect to $M/200$, for H1 (top) and CE (bottom). Normalized bias were calculated using Eq.~\eqref{eqn:NB}.}
    \label{tab:q1_H1_bias}
\end{table*}

It has been shown that waveforms involving significant contributions from higher modes \cite{PhysRevD.104.044037}, require a higher resolution to satisfy accuracy requirements in comparison to waveforms that lack significant contributions from higher modes.
As such we also injected the same $q =1$ NR waveforms but at $\iota = \pi/6$. At this inclination, the detected GW is shaped not just by the modes observed at $\iota=0$ but also by $(2, -2)$ and $(4, 4)$. To quantify the impact of truncation errors as a function of SNR, the SNR selection was done the same way as for the face-on case.
For both of these cases, we use $M_\text{tot} = 100 M_\odot$ to ensure the majority of the injected waveform falls in band. 

We also extended our study to include $q=1/3$ waveform injections which were generated at $M_\text{tot} = 150 M_\odot$ and $\iota=0$. 
We adjusted our SNR selection approach slightly, opting for six SNRs instead of eight. 
This selection involved choosing six resolution levels in such a way that, for the first three levels, two corresponded to scenarios where $M/100$ is predicted to be indistinguishable, and one where it is predicted to be distinguishable. 
We applied the same methodology to determine the next three SNRs, this time focusing on $M/120$. 
Since the $q=1/3$ injections were generated at a different total mass, to ensure the injected waveform starts in band,
it would be unreasonable to directly compare the results of $q = 1$ and $q = 1/3$ injections. To enable a direct comparison, the same total mass must be injected for both  $q = 1$ and $q = 1/3$ cases. In such a comparison, we would expect that in the $q = 1/3$ case, the same resolution introduces a significantly higher parameter bias than in the $q = 1$ case, at a given SNR. To emphasize this point, we injected a set of $q=1$ waveforms at $M_\text{tot} = 150 M_\odot$, detailed in Sec.~\ref{ssec:q3}.

Furthermore, it is important to note that parameter biases are also influenced by the shape of the noise power spectral density (PSD) of the detector. Therefore, our study encompassed two detectors, namely design aLIGO \cite{PSD-1,PSD-2} and CE.
Figure~\ref{fig:ch_strain} depicts the characteristic strain~\cite{Moore:2014lga} for the face-on $q=1$ and $q=1/3$ waveforms, alongside the noise amplitude spectral densities for comparison.

For each set of NR injections, differing only in resolution and recovered at a sequence of SNRs, we kept our settings the same so any differences in the posterior can be attributed to the difference in NR resolution. 
However, despite our best efforts, there are other errors that can potentially impact the final posterior distribution, with some of them being Monte Carlo integration errors when evaluating marginalized likelihood and the fit to the finite grid of intrinsic parameters.

For our PE runs, in the first iterative step of {\tt RIFT}, where the marginalized likelihood is evaluated, we marginalized over all the extrinsic parameters, except for ra and dec which remained fixed for all runs. Thus we marginalized over $t_c$, $\phi_c$, $\psi$, $D_L$, and $\iota$ with the likelihood integration starting from $20$ Hz and ending at $2048$ Hz, for both H1 and CE. We chose to maintain the same frequency range, for both H1 and CE, to ensure a direct comparison of PE results.  In the second iterative step, we approximated the likelihood using random forests, and the Monte Carlo sampling was carried out using Gaussian mixture models. To ensure our choice of likelihood approximation and sampling method used within {\tt RIFT} does not impact our final results, we reanalyzed the highest SNR injection for each synthetic source using Gaussian process regression as the approximation method and adaptive cartesian as the sampling method (refer to \cite{Lange_2017,Wofford:2022ykb} for details on the different approximation and sampling method implemented in {\tt RIFT}).
We found our results remained consistent regardless of our choice of likelihood approximation and sampling method.

\begin{figure}[tb]
    \includegraphics[width = \columnwidth]{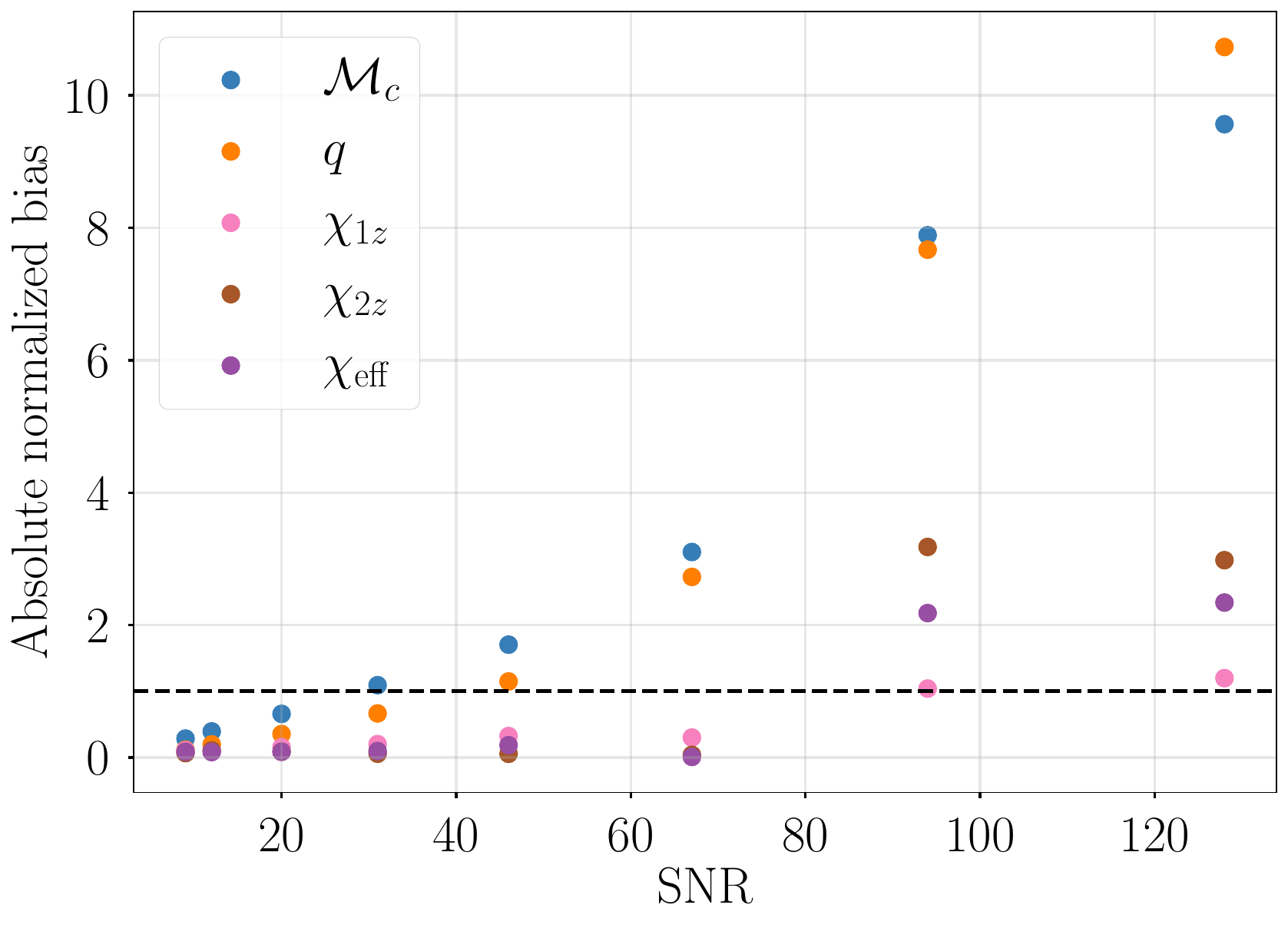}
    \caption{\textbf{Absolute normalized bias in multiple parameters}: Plot showing absolute normalized bias in 1D marginal posteriors for selected source parameters when $q=1$, $\Delta =  M/80$ is used as an injection.  The dashed horizontal line marks an absolute normalized bias of unity.}
    \label{fig:param_bias}
\end{figure}

To quantify the impact of truncation errors on PE, we calculate normalized bias for all parameters. The normalized bias for a parameter $\lambda$ is calculated as:
\begin{equation}
    \text{Normalized bias}= \Delta \tilde{\lambda}/\sigma
    \label{eqn:NB}
\end{equation}
where $\tilde{\lambda}$ is the median of the marginalized $\lambda$ posterior distribution, $\Delta \tilde{\lambda}$ is the shift in the median of marginalized $\lambda$ posterior distribution relative to the highest resolution posterior and $\sigma$ is the standard deviation of the highest resolution posterior. In Fig.~\ref{fig:param_bias}, we plot normalized bias in multiple parameters when the $q=1, \iota=0, \Delta = M/80$ waveform is used as injection. We observe that $\mc_c$ is the first parameter to display an absolute normalized bias of unity, which we will consider as a significant parameter bias in our work. Therefore, moving forward, we place primary emphasis on $\mc_c$.

\subsection{$\boldsymbol{q=1}$}
\label{ssec:q1}

For the $q=1$ system, we generated injections from each of the four differently resolved waveforms at eight different SNRs.  We started from SNR $9$, where $M/70$ is predicted to be the critical grid spacing, and went all the way up to SNR $128$, where $M/135$ is predicted to be the critical grid spacing. All eight sets of injections had identical parameters except $D_L$.

\begin{figure*}
    \includegraphics[width=\columnwidth]{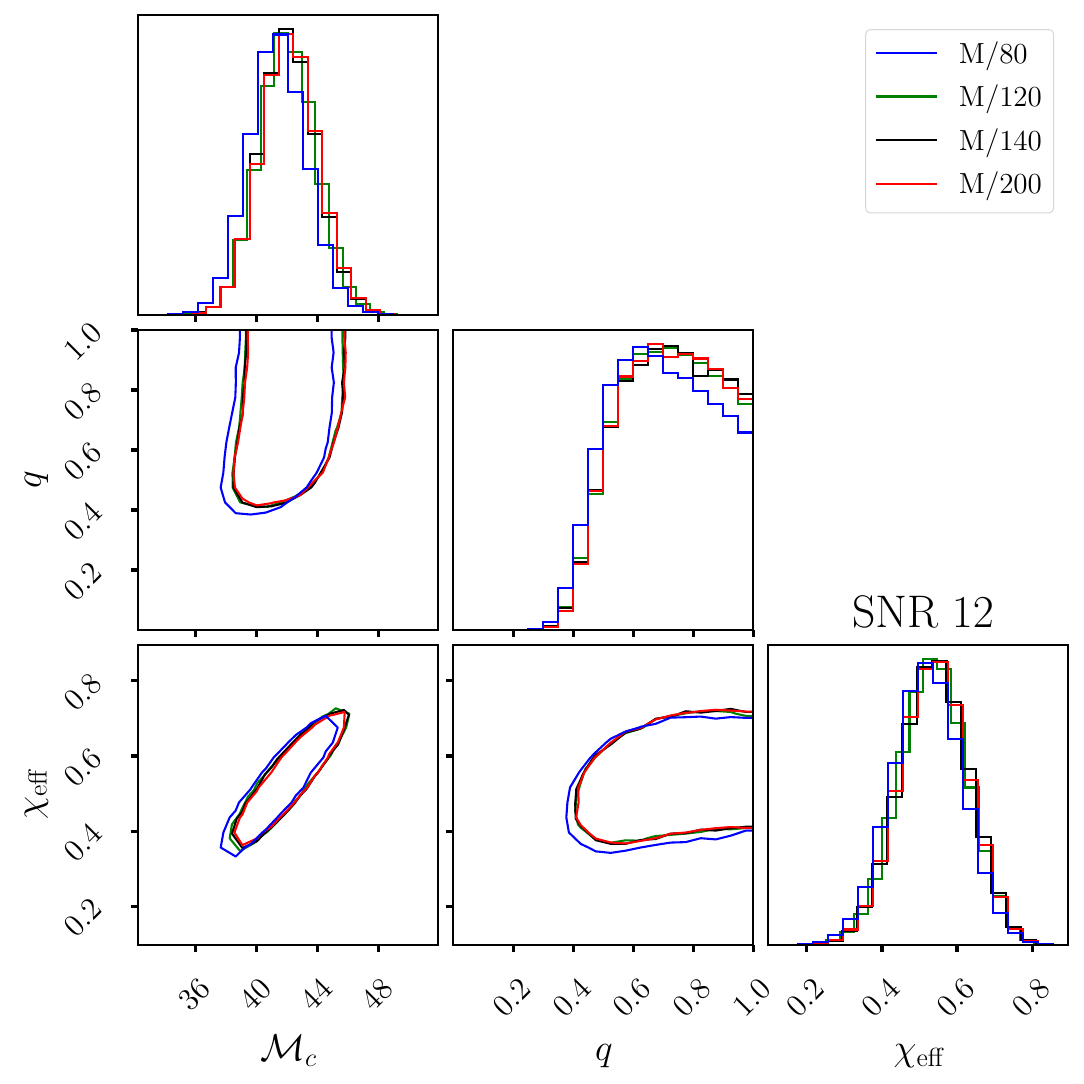}
    \includegraphics[width=\columnwidth]{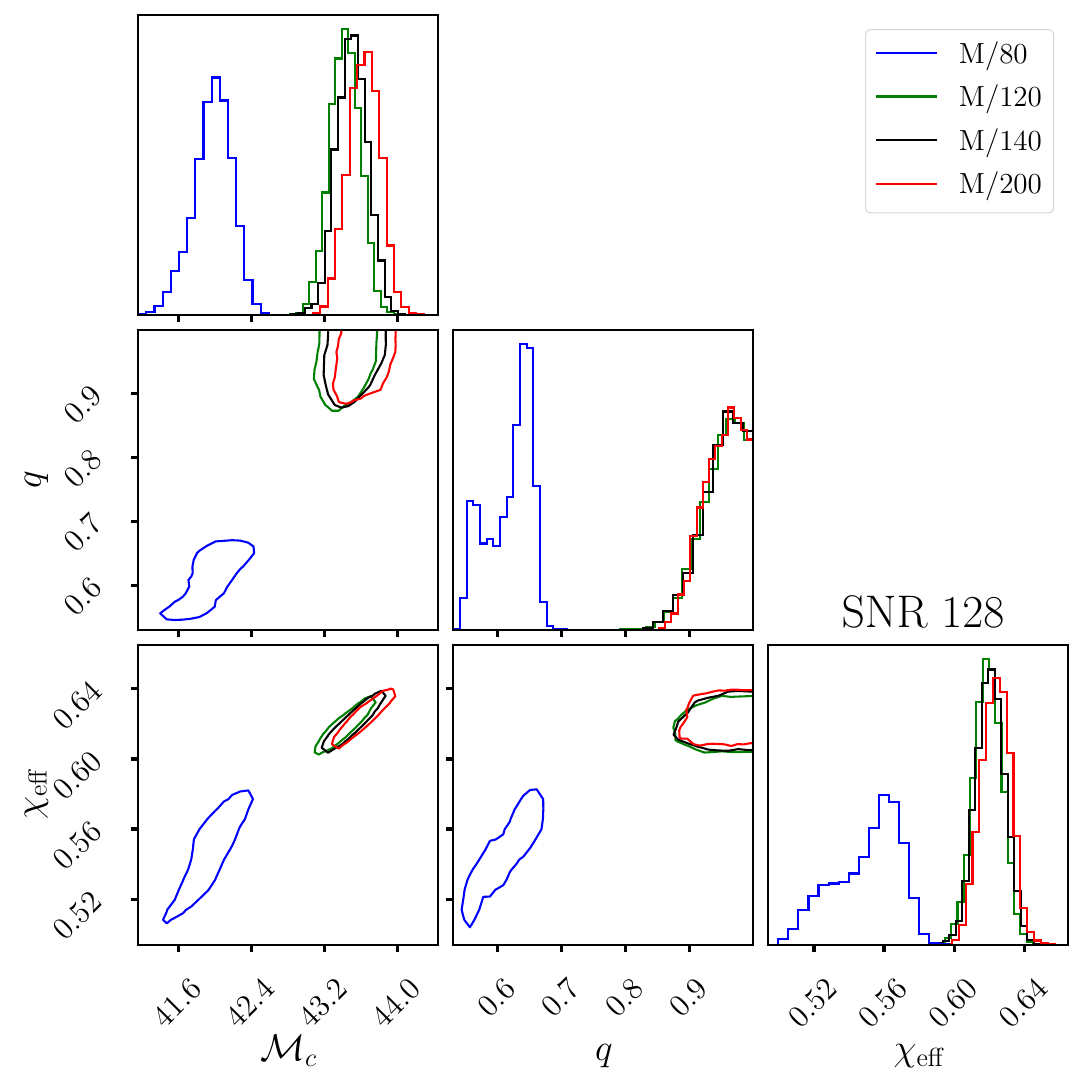}
    \includegraphics[scale = 0.5] {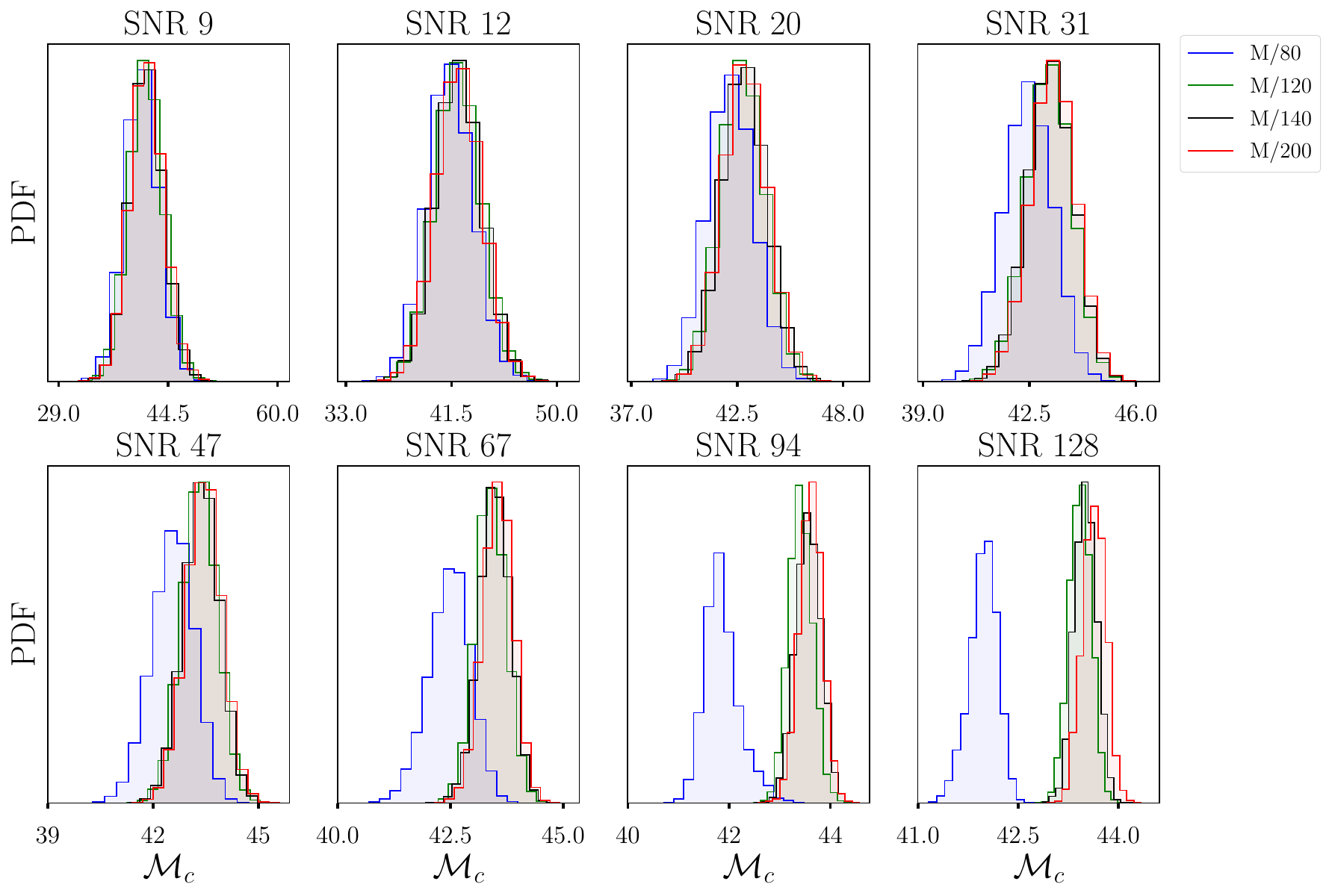}
    \caption{\textbf{PE results for $\boldsymbol{q=1, \iota = 0}$ injections (H1)}:  \textit{Top left}: One- and two-dimensional marginal posterior distributions for $\mathcal{M}_c, q,$ and $\chi_{\text{eff}}$. Diagonal panels show the one-dimensional marginal posterior distribution, while contours in the off-diagonal panels show the 90\% credible intervals for the two-dimensional marginal posterior distribution. Different colored curves correspond to different resolutions. Injections had an SNR of $12$ and the minimum resolution for indistinguishability at that SNR is predicted to be $(M/75)^{-1}$. \textit{Top right:} Corner plot produced after performing PE at an SNR of $128$, where the minimum resolution for indistinguishability is predicted to be $(M/135)^{-1}$.
    \textit{Bottom}: One-dimensional marginalized posterior distributions for $\mc_c$ are presented here. PE was conducted at a sequence of SNRs, with all parameters held constant except for $D_L$. Each panel illustrates the outcomes for a specific SNR, and distinct colored curves represent different resolutions. With increasing SNR, both $M/80$ and $M/120$ posteriors gradually separate from the others.}
    \label{fig:corner_H1_q1}
\end{figure*}

The top panel of Fig.~\ref{fig:corner_H1_q1} shows selected 1-D and 2-D marginals of our recovered posteriors for SNR $12$ and SNR $128$, where $M/75$ and $M/135$ are the predicted critical grid spacings respectively. 
At SNR $12$, the 1-D and 2-D posterior distributions for all four resolutions lie on top of each other, whereas at SNR $128$ the $M/80$ posteriors peel away from the rest, introducing notable biases in all three parameters displayed. 
Consequently, at SNR $12$, all four resolutions yield equivalent results for PE, rendering the use of higher-resolution waveforms unnecessary. However, at SNR $128$, the use of $M/80$ leads to biased PE. Additionally, we observe truncation errors cause a downward shift in the marginalized posteriors of the three parameters relative to the highest resolution posterior.

The bottom panel of  Fig.~\ref{fig:corner_H1_q1} shows the detector-frame $\mathcal{M}_c$ posterior distribution for all eight SNRs. 
The posteriors beyond SNR 12, where the influence of prior becomes negligible, peak at the same point indicating that the shift in the lower resolution $\mc_c$ posteriors relative to the highest resolution posterior remains consistent; only its significance increases with rising SNR.
From Table \ref{tab:q1_H1_bias}, we can see that $M/80$ produces an absolute normalized bias greater than one at  SNR $29$, suggesting that $M/80$ might be accurate for SNRs lower than $29$ but beyond this SNR, it is no longer sufficiently accurate for unbiased PE. 
Also, even though $M/80$ becomes distinguishable at SNR $15.7$, according to the estimate of Eq.~\eqref{eqn:criteria}, it does not produce a normalized bias of one, underscoring the fact that the criterion of indistinguishability arising from matches does not imply significant parameter biases in PE.

\begin{figure*}
    \includegraphics[width=\columnwidth]{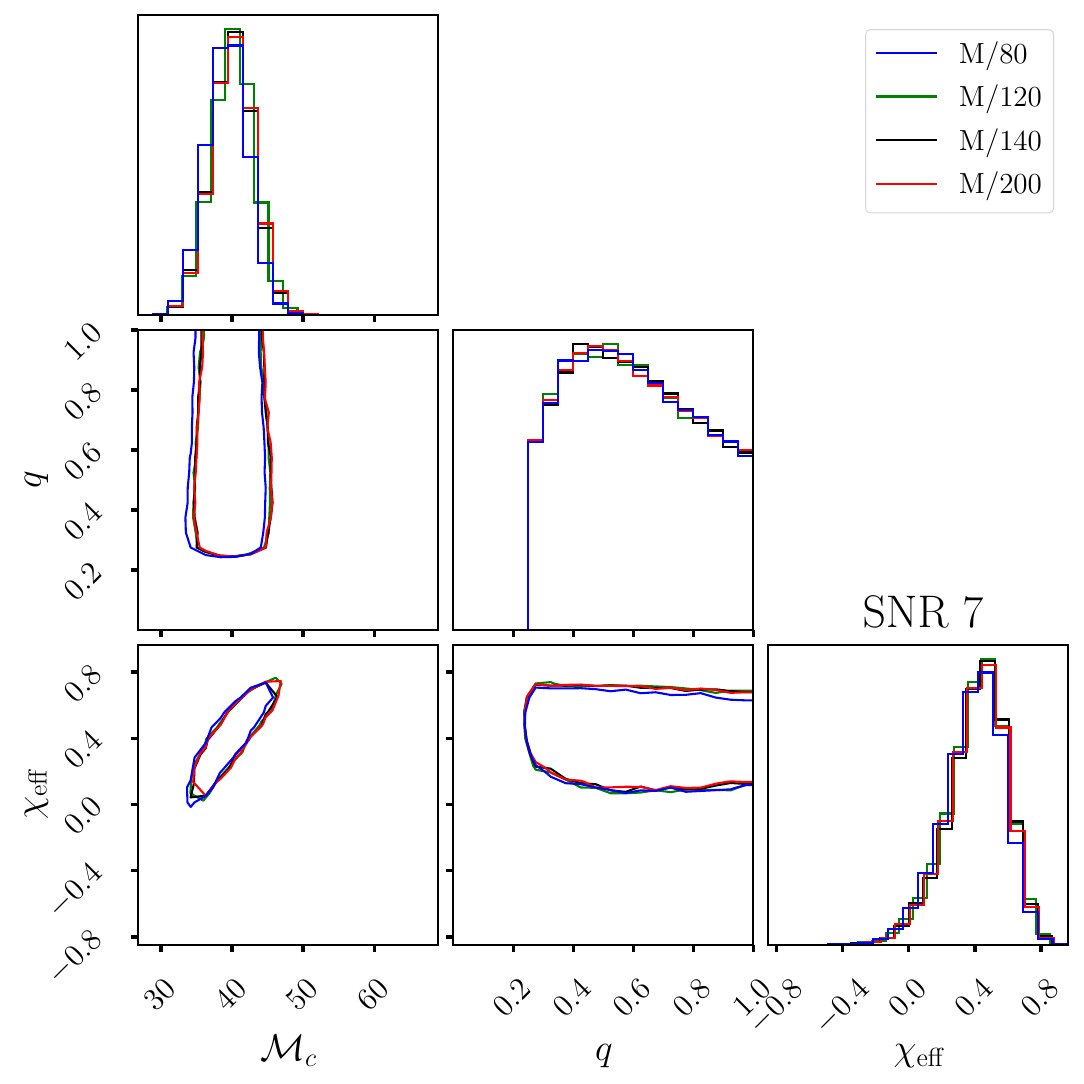}
    \includegraphics[width=\columnwidth]{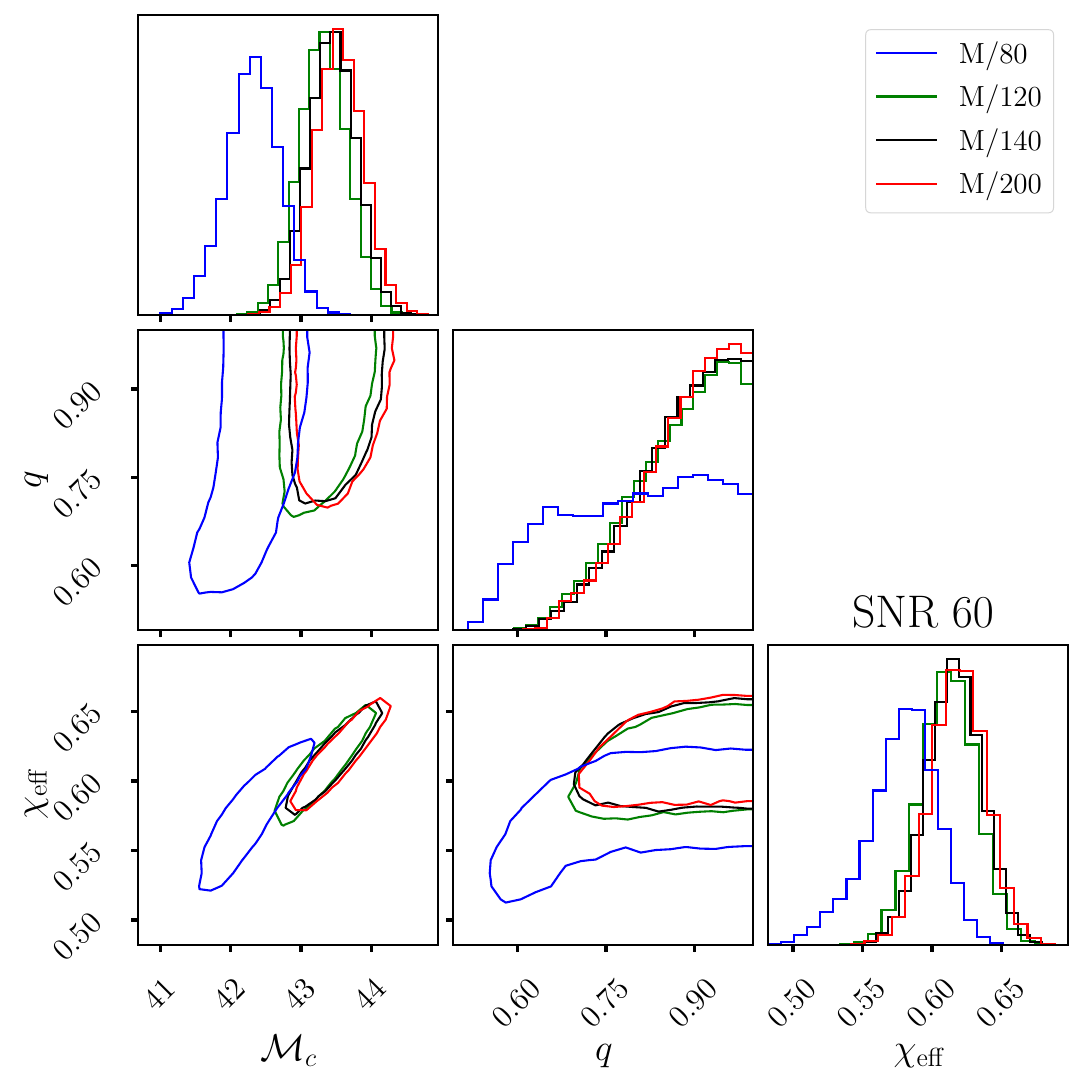}
    \includegraphics[scale = 0.5]{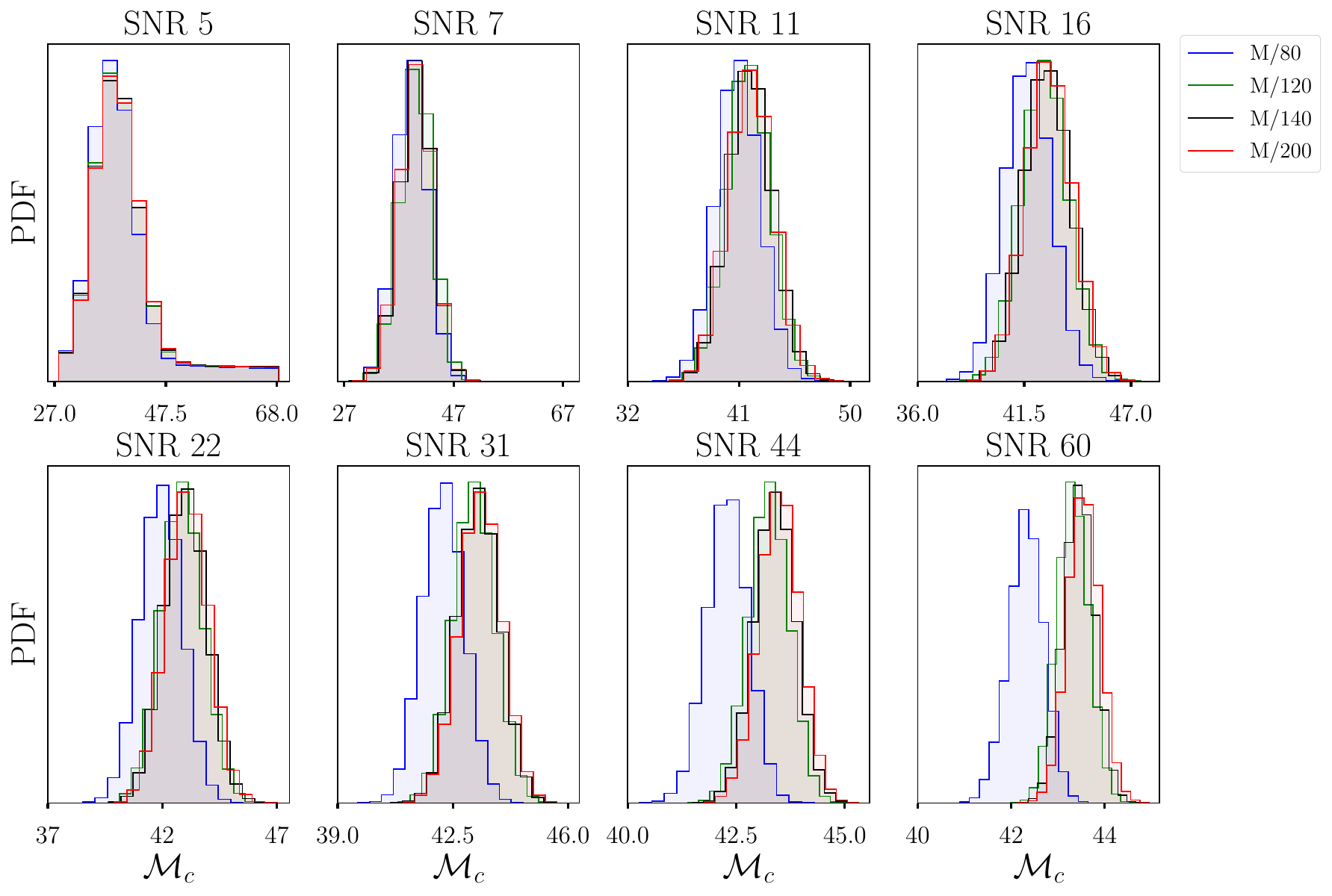}
    \caption{\textbf{PE results for $\boldsymbol{q=1, \iota = 0}$ injections (CE)}:  \textit{Top left}: One- and two-dimensional marginal posterior distributions for $\mathcal{M}_c, q,$ and $\chi_{\text{eff}}$. Diagonal panels show the one-dimensional marginal posterior distribution, while contours in the off-diagonal panels show the 90\% credible intervals for the two-dimensional marginal posterior distribution. Different colored curves correspond to different resolutions. Injections had an SNR of $7$ and the minimum resolution for indistinguishability at that SNR is predicted to be $(M/75)^{-1}$. \textit{Top right:} Corner plot produced after performing PE at an SNR of $60$, where the minimum resolution for indistinguishability is predicted to be $(M/135)^{-1}$.
    \textit{Bottom}: One-dimensional marginalized posterior distributions for $\mc_c$ are presented here. PE was conducted at a sequence of SNRs, with all parameters held constant except for $D_L$. Each panel illustrates the outcomes for a specific SNR, and distinct colored curves represent different resolutions. With increasing SNR, the $M/80$ posterior gradually separates from the others. }
    \label{fig:corner_CE_q1}
\end{figure*}

We then repeated our analysis for CE, which has a different PSD shape than that of H1. Examining Fig.~\ref{fig:ch_strain}, we notice that H1 and CE exhibit a similar shape at higher frequencies. 
However, at lower frequencies, CE surpasses H1 in terms of sensitivity, which leads to greater mismatches and subsequently reduces the critical SNR for a given resolution. 
Similar to our H1 analysis, we selected eight SNRs, ranging from $5$ to $60$, where the predicted critical grid spacings are $M/70$ and $M/135$ respectively. 
The results of PE are shown in Fig.~\ref{fig:corner_CE_q1}; the upper part displays the 1-D and 2-D marginalized posteriors for SNR $7$ and SNR $60$, corresponding to critical grid spacings of $M/75$ and $M/135$. 
At SNR $7$, the 1-D and 2-D posterior distributions for all four resolutions closely overlap. 
However, at SNR $60$ the $M/80$ posteriors deviate notably from the others, introducing significant biases in all three parameters under consideration. 
The lower part of Fig.~\ref{fig:corner_CE_q1} presents the $\mc_c$ posterior distribution for all eight SNRs. 
We observe that $M/80$ introduces significant biases at an SNR of $18.3$, which is considerably lower than what we observed for H1. 
This is due to the differences in the shape of the noise PSD of H1 and CE. 
Therefore, it is necessary to re-evaluate the accuracy of NR waveforms when the shape of the PSD undergoes alterations.

\begin{figure}
    \centering
    \includegraphics[width = \columnwidth] {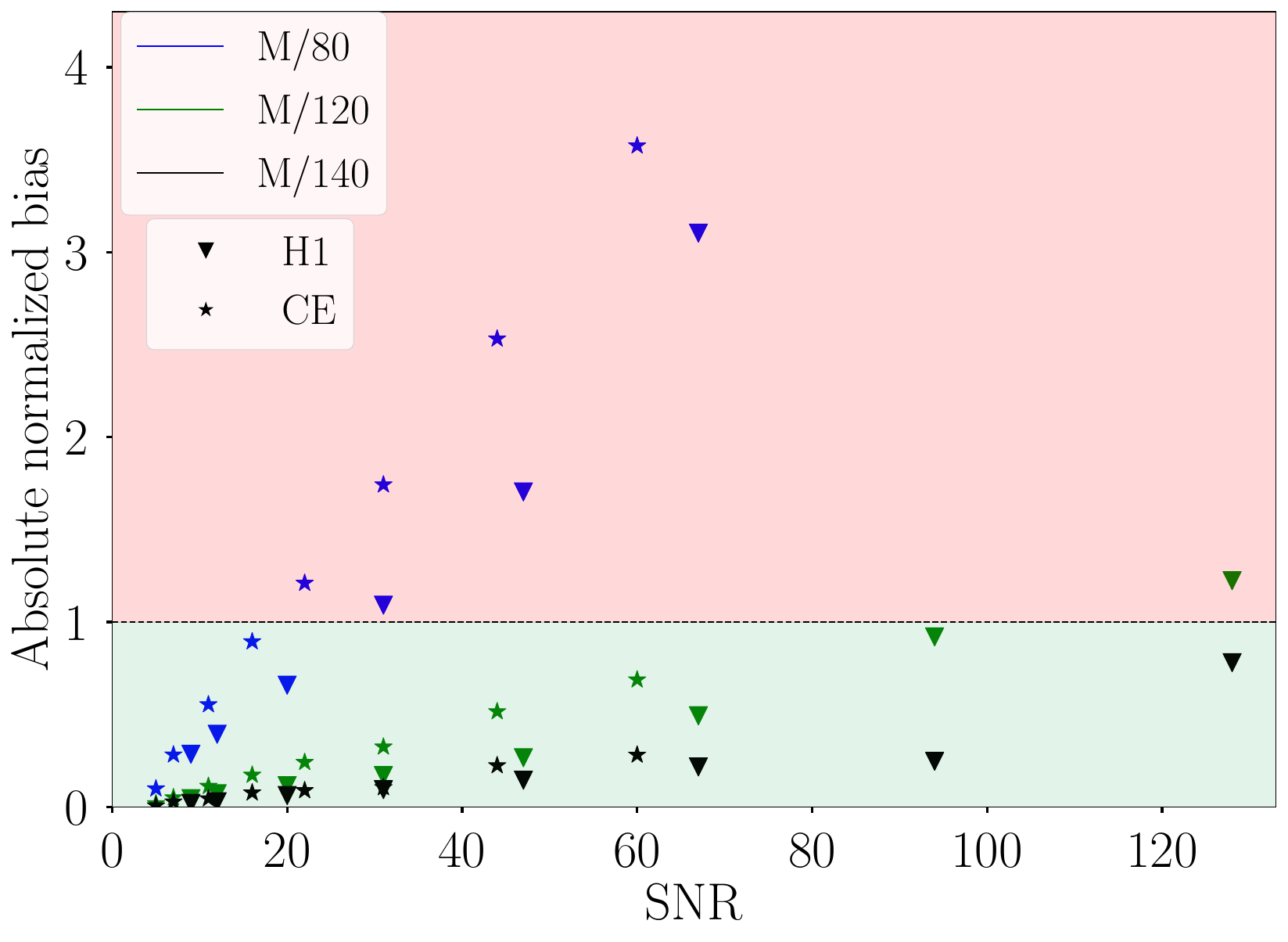}
    \caption{\textbf{Absolute normalized bias in the marginalized $\mathcal{M}_c$ posterior distributions of $\boldsymbol{q=1, \iota = 0}$ injections as a function of SNR}: The dashed horizontal line marks an absolute normalized bias of unity.  We have omitted normalized bias values exceeding five to concentrate on the region where the bias is around one. The bias values were calculated with respect to $M/200$.}
    \label{fig:bias_q1}
\end{figure}

We summarize these results in Fig.~\ref{fig:bias_q1}, which shows the absolute normalized bias in $\mc_c$ as a function of SNR across injection resolutions and detector PSDs. 
It is evident that only in the case of the lowest grid spacing, $M/80$, the median of $\mc_c$ exhibits a bias surpassing the statistical uncertainty in parameter recovery for both PSDs, and this occurs only at SNRs above approximately $18$. Additionally, at the highest explored SNR, $M/120$ exhibits a significant bias in H1.

\subsection{$\boldsymbol{q=1}, \boldsymbol{\iota = \pi/6}$}
\label{ssec:q1inc}
Higher GW modes require more resolution to be fully resolved than the dominant $(2,2)$ mode and as such the same resolution will produce more parameter bias at non zero inclination, due to greater contribution from higher modes, than it will for zero inclination. To illustrate this, we injected the same $q=1$ waveforms but at a modest inclination of $\pi/6$. 
Even at such a small inclination, two additional GW modes, $(2, -2)$ and $(4, 4)$, significantly contribute to the detected GW, thereby increasing the mismatch and consequently decreasing the critical SNR for the same resolution. To estimate the impact of bias as a function of SNR, we repeat the analysis the same way we did for the face-on case.
We started from SNR $6$ ($4$), where $M/70$ is the predicted critical grid spacing, and went all the way up to SNR $79$ ($55$), where $M/135$ is the predicted critical grid spacing for H1 (CE). The results for this case are broadly the same as for the $q=1$, $\iota = 0$ case.
Only the lowest resolution $(M/80)^{-1}$ injection displays significant bias in its recovered parameters, and truncation errors continue to cause a downward shift in the marginalized posteriors of the three parameters relative to the highest resolution posterior.
We provide our posteriors and absolute normalized bias values in Appendix~\ref{assec:q1_incl30}.

We observe an absolute normalized bias of unity due to $M/80$ at an SNR of $28.5$ ($18.1$) for H1 (CE), lower than $\iota = 0$ case, showing that the accuracy requirements change as the observed inclination changes. 
For a more direct comparison, we also recovered inclined injections at SNR $128$ ($60$) for H1 (CE). At these SNRs, we found that $M/80$ produces an absolute normalized bias of $10.72$ ($3.63$), surpassing the bias observed at the same SNR for the face-on case which was $9.59$ ($3.58$) for H1 (CE).

\subsection{$\boldsymbol{q=1/3}$, $\boldsymbol{\iota = 0}$}
\label{ssec:q3}

A resolution that is considered adequate for unbiased PE for $q = 1$ systems may prove insufficient for systems with unequal mass ratios. This is due to the fact that as the mass ratio decreases, a higher resolution is needed to accurately resolve the smaller black hole. Additionally, the inherent asymmetry in the system induces the excitation of higher modes, which require more resolution to be sufficiently resolved. 
The combination of these factors leads to an increase in the parameter bias introduced by a resolution as the mass ratio decreases. To illustrate this, we extended our analysis to include $q=1/3$ waveforms. The waveforms were extracted from simulations that had an initial separation of $9M$, compared to the initial separation of $12M$ for $q=1$ simulations, resulting in shorter waveforms. 
To ensure that the waveforms do not begin in the frequency band of interest, we increased $M_{\text{tot}}$ from $100 M_\odot$ to $150 M_\odot$. As such, we cannot directly compare the normalized bias values for these injections and for $q=1$ injections.


\begin{figure}
    \centering
    \includegraphics[width = \columnwidth] {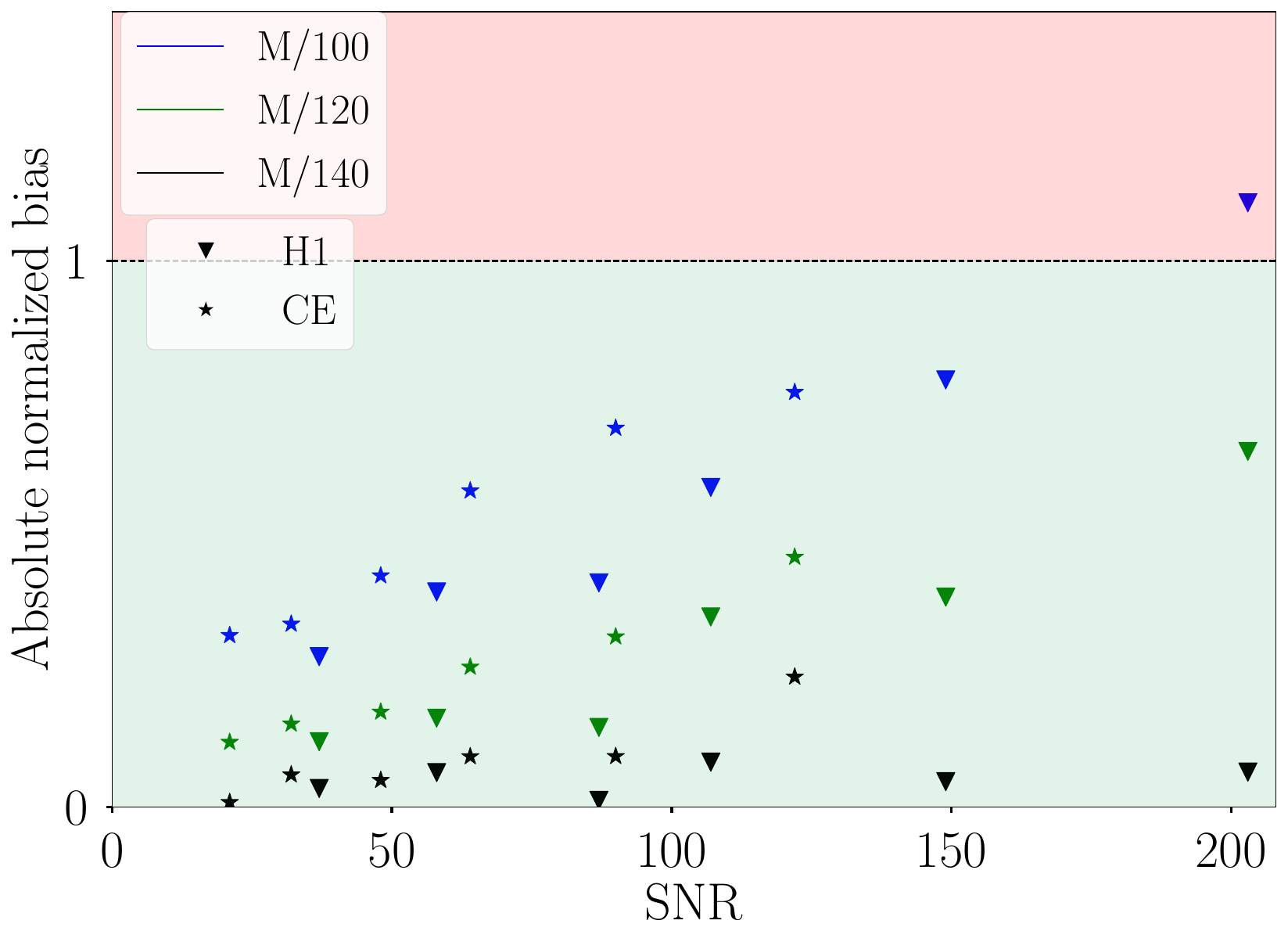}
     \caption{\textbf{Absolute normalized bias in the marginalized $\mathcal{M}_c$ posterior distributions of $\boldsymbol{q=1/3, \iota = 0}$ injections as a function of SNR}: The dashed horizontal line marks an absolute normalized bias of unity. The bias values were calculated with respect to $M/180$.}
    \label{fig:bias_q3}
\end{figure}

The analysis was carried out the same way as we did for $q = 1$ injections. We started at SNR $37$ ($21$) and went all the way to SNR $203$ ($122$) for H1 (CE), to study the impact of bias as a function of SNR. We observe that $M/100$ produces a normalized bias of unity at $185.4$ ($117.1$) for H1 (CE). The PE results for this system are summarized in Fig.~\ref{fig:bias_q3}, reflecting similar qualitative trends observed for the previous two synthetic sources. The posteriors and absolute normalized bias values are provided in Appendix~\ref{assec:q3}. 

To emphasize the crucial point that the parameter bias introduced by a resolution increases as the mass ratio decreases, we injected $q=1$ waveforms at $M_\text{tot} =150 M_\odot$ at SNR $203$ ($122$) for H1 (CE). We find that $(M/120)^{-1}$ resolution introduces an absolute normalized bias of $0.31$ ($0.16$) for $q=1$ and $0.65$ ($0.46$) for $q=1/3$ for H1 (CE). This shows that at a given SNR and NR resolution, waveforms with lower mass ratios introduce greater bias compared to those with higher mass ratios, necessitating a more stringent accuracy requirement.

\section{Extrapolation}
\label{sec:extrapolation}

In Sec.~\ref{sec:PEresults}, we assessed the errors introduced in PE when finite resolution NR waveforms are used.
We accomplished this by comparing the PE recovery of low-resolution NR injections with those of the highest-resolution NR injection, for each of the three synthetic sources. 
However, it is important to acknowledge that any finite-resolution waveform can introduce bias into PE, and this bias persists even when we use the highest-resolution waveform. 
As such, in this section, we employ an extrapolation procedure to predict the median of a marginalized posterior distribution for parameter $\lambda$, recovered for a hypothetical $M/\infty$ injection. We then use this extrapolated value to calculate the bias in PE attributable to the use of the highest-resolution waveforms.

We utilize the understanding that the bias in NR waveforms is proportional to $\Delta^\alpha$. 
Thus, if all aspects of PE remain consistent across the recovery of each differently resolved injection, the bias in the final posterior due to finite resolution should also be proportional to $\Delta^\alpha$. 
Keeping this in mind, we fit the following function to the posteriors recovered for all four injections:
\begin{equation}
    \lambda = a\Delta^{\alpha} + b \,.
    \label{eq:fit}
\end{equation}
Here $a$ and $b$ are fitting constants, and $\alpha$ takes a value of $4$ \cite{PhysRevD.104.044037} for the NR waveforms used in this study. 
We then fit this function to the $\mc_c$ posteriors, for all three systems and for both detectors at the SNRs mentioned in Table \ref{tab:extrapolation_normalized}. 
Analyzing the posteriors at the SNRs in Table \ref{tab:extrapolation_normalized}, we find the standard deviation tends to remain approximately the same for all the different resolution $\mc_c$ posteriors. Using this observation, alongside the median of  $\mc_c$ posterior for $M/\infty$ injection, we can determine the normalized bias in the highest resolution $\mc_c$ posterior, which are listed in Table \ref{tab:extrapolation_normalized}. 
Our findings show that the highest-resolution waveform utilized for each synthetic source meets the accuracy requirements for the injected SNRs, validating the comparison of low-resolution posteriors with the highest-resolution posterior.

\begin{figure}
    \centering
    \includegraphics[width = \columnwidth] {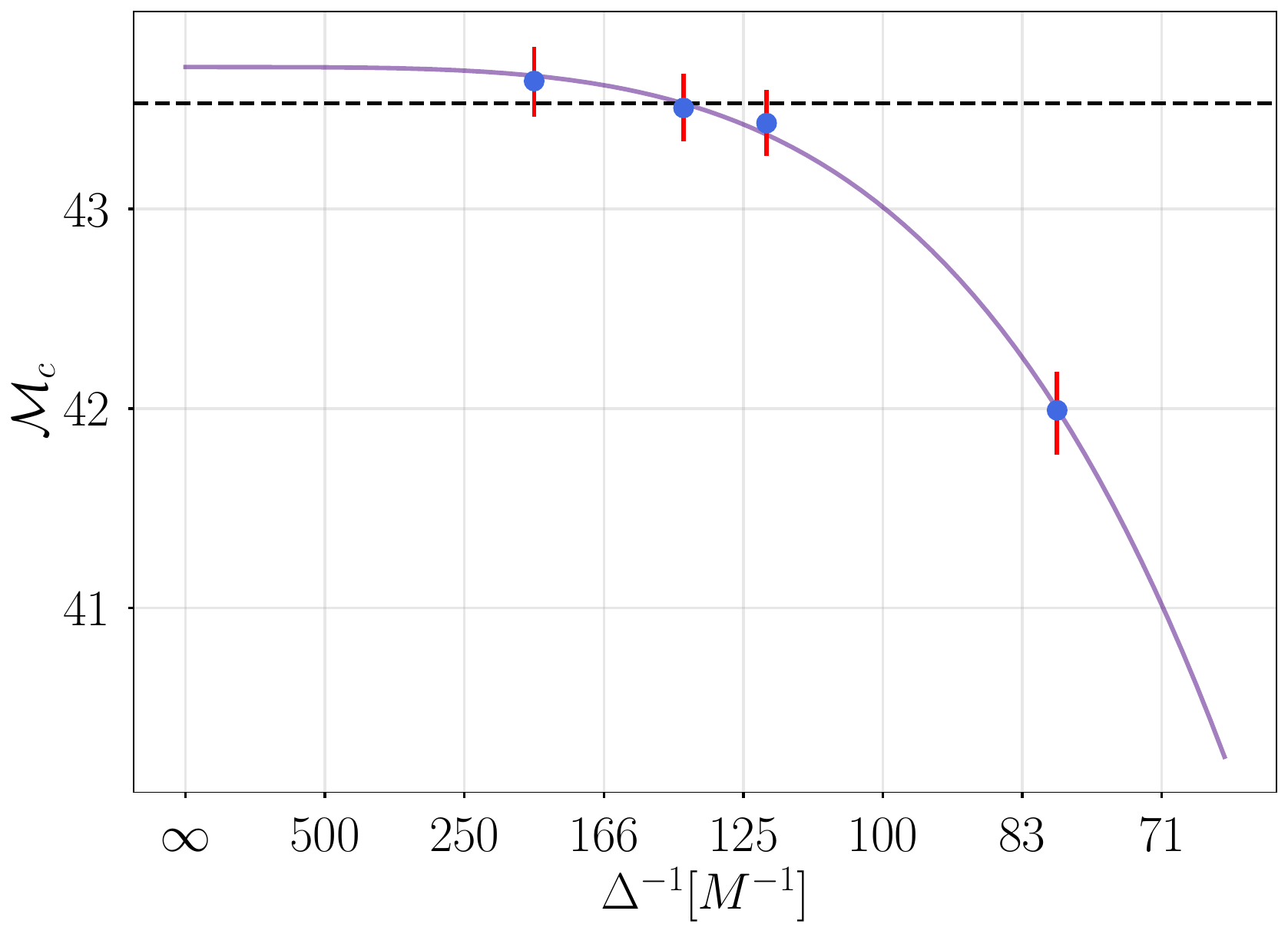}
    \caption{\textbf{Extrapolation procedure}: This figure shows our extrapolation procedure applied to the recovered marginalized $\mc_c$ posterior for all four H1 $q=1, \iota = 0$ injections at SNR $128$. We employ the fitted curve to determine the recovered $\mc_c$ for a hypothetical $M/\infty$ injection and use it to calculate the bias in our highest resolution injection posterior. The data points represent the median values of recovered $M_c$ distributions, the error bars represent one standard deviation and the indigo curve illustrates the fitted curve. The horizontal dashed line represents the injected value.}
    \label{fig:fit}
\end{figure}

Figure \ref{fig:fit} demonstrates the application of the extrapolation procedure. It is clear from the figure that the $\mc_c$ median at $M/\infty$ does not align with the injected value. 
This discrepancy is expected, due to differences between the extrapolated NR waveforms $h_\infty$ and the recovery model \NRSur $h_{\rm sur}$, as discussed in Sec.~\ref{ssec:Approach}. 
We find that the differences between $h_\infty$ and $h_{\rm sur}$ is ultimately due to mismatches between the {\tt MAYA} waveforms we use for our injections, and the {\tt SpEC} \cite{Boyle_2019} waveforms which underlie $h_{\rm sur}$.
For the parameters shows in Table~\ref{tab:param}, the mismatches between {\tt SpEC} and $h_{\rm sur}$ are on the order of $10^{-4}$. 
Meanwhile, for the same parameters, the mismatches between {\tt MAYA} and $h_{\rm sur}$ and between {\tt MAYA} and {\tt SpEC} waveforms are both on the order of $10^{-3}$, 
suggesting that the observed parameter discrepancy can be ascribed to differences between {\tt SpEC} and {\tt MAYA} waveforms. 
These differences can be possibly due to the fact that the {\tt MAYA} waveforms used in our study were extracted at finite radius whereas the {\tt SpEC} waveforms used in the construction of the recovery model were extrapolated to infinity. 
Furthermore, the two NR codes employ different methods to solve partial differential equations, which could also explain the discrepancy.

\begin{table}
\centering
\addtolength{\tabcolsep}{4pt} 
     \begin{tabular}{c c c c}
     \hline
     System & $\Delta$ & H1 & CE \\ [0.5ex] 
     \hline\hline
    $q=1$ & $M/200$ & 0.59 (128) &  0.082 (60)\\ 
    $q=1, \iota = \pi/6$ & $M/200$ &  0.87 (128) & 0.092 (60)\\
    $q=1/3$ & $M/180$ & 0.15 (203) & 0.005 (122)\\[1ex] 
     \hline
     \end{tabular}
     \caption{\textbf{Absolute normalized bias in the highest resolution posteriors}: Numbers in parenthesis are the SNRs at which these biases are evaluated.}
    \label{tab:extrapolation_normalized}
\addtolength{\tabcolsep}{-4pt} 
\end{table}

\section{Prediction}
\label{sec:prediction}

\begin{table*}
\centering
\addtolength{\tabcolsep}{4pt} 
     \begin{tabular}{c c c c c c c}
     \hline
      System & $\Delta$  & Detector & Critical SNR  & 
      Critical SNR      & Critical SNR   & Critical SNR \\
      &  &  & from PE & [Eq.~\eqref{eqn:criteria}] & [Eq.~\eqref{eq:D_criterion}]& [Eq.~\eqref{eqn:mod_criteria}] \\ [0.5ex] 
     \hline\hline
     \vspace{0.10cm}
    \multirow{2}{*}{$q=1$}& \multirow{2}{*}{$M/80$}&H1 & 29.0 & 15.7 & 32.3 & 31.5\\ 
    & & CE & 18.3 & 8.3 & 17.0 & 16.6\\ 
    \hline
    \vspace{0.10cm}
    \multirow{2}{*}{$q=1, \iota = \pi/6$}&\multirow{2}{*}{$M/80$}& H1 &  28.5 & 15.5 & 31.9 & 31.0\\ 
    & & CE & 18.1 &  8.2 & 17.0 & 16.5 \\ 
    \hline
    \vspace{0.10cm}
    \multirow{2}{*}{$q=1/3$}& \multirow{2}{*}{$M/100$}&H1 & 185.4 & 71.2 &  157.4& 142.3 \\ 
    & & CE & 117.1 & 39.9 & 88.0& 79.7\\ 
     \hline
     \end{tabular}
     \caption{\textbf{Comparing the critical SNR from full PE, Eq.~\eqref{eqn:criteria}, Eq.~\eqref{eq:D_criterion}, and Eq.~\eqref{eqn:mod_criteria}:}  The critical SNRs predicted using the modified criterion (Eq.~\eqref{eqn:mod_criteria}) tend to be on the conservative side.}
    \label{tab:comparison_criteria}
\addtolength{\tabcolsep}{-4pt} 
\end{table*}

In this section, we aim to determine the SNR at which a finite-resolution NR waveform will produce an absolute normalized bias of unity in at least one of the parameters, without having to do full Bayesian inference. 
While current PE codes have undergone significant speedups \cite{Wysocki_2019,Smith_2020}, full Bayesian inference can still be computationally intensive, especially for signals with higher SNRs. Furthermore, when determining the minimum resolution required for an NR waveform to avoid significant bias in GW interpretation for a particular detector, full PE may be excessive.  
Therefore we employ the following criterion \cite{Chatziioannou:2017tdw,PhysRevResearch.2.023151} to give an approximate estimation of critical SNR:
\begin{equation}
    \epsilon[h_1,h_2] < D/2\rho^2 \,.
    \label{eq:D_criterion}
\end{equation}
Here, the pre-factor $D$ is the number of intrinsic parameters affected by waveform inaccuracy. 
Our study was conducted on aligned systems, and as such we have set $D$ equal to $4$. 
To get a more accurate assessment of critical SNR, one would need to tune $D$ by finding the exact SNR at which statistical error becomes equal to systematic error. 
However, such a refinement typically requires the application of PE and as such we rely on an approximate value of 4 for $D$.  
Examining Table~\ref{tab:comparison_criteria}, we observe a reasonably good agreement between the critical SNR values obtained through full PE and those determined by the criterion of Eq.~\eqref{eq:D_criterion} for both $q=1$ and $q=1/3$ systems. 
The disparity between the two SNR values is approximately $11\%$ for $q = 1$ systems and roughly $16\%$ for $ q = 1/3$ systems. 
It is worth noting that, with the exception of $q = 1$ systems as observed by H1, Eq.~\eqref{eq:D_criterion} tends to provide conservative estimates for the critical SNR. 

While Eq.~\eqref{eq:D_criterion} is relatively straightforward to apply, it has a few notable limitations when it comes to its application to finite-resolution NR waveforms. 
To determine the minimum resolution needed for unbiased PE for a detector, 
one would generate NR waveforms at various resolutions and then identify the resolution with a critical SNR exceeding the SNR of interest for that detector. 
Furthermore, mismatches would need to be calculated using an infinite-resolution waveform, which is not feasible. 
To address these limitations, we adjust Eq.~\eqref{eqn:criteria} by introducing the pre-factor $D$. 
This modification renders the equation more realistic in its estimation of the required SNR as it takes into account the dimensionality of the search parameter space. 
By utilizing this adjusted criterion, one can determine the minimum resolution necessary to ensure that PE remains largely unaffected by finite resolution waveforms, all without the need to generate waveforms for a sequence of resolutions or an infinite resolution waveform. 
One would just need to generate waveforms at two resolutions, compute $\beta$, and then find the resolution needed using the modified criterion. 
The modified criterion is:
\begin{equation}
    (\beta \Delta^\alpha)^2 < D/\rho^2 \,.
    \label{eqn:mod_criteria}
\end{equation}
Here $\beta$ is calculated using Eq.~\eqref{eq:beta}, $\alpha$ is the convergence rate and $\Delta^{-1}$ is the resolution. 
From Table~\ref{tab:comparison_criteria}, we can see that the predictions obtained through the modified criterion closely align with those from  Eq.~\eqref{eq:D_criterion}. 
The difference in the critical SNR values by full PE and predictions by modified criterion is approximately $9\%$ for $q = 1$ systems and roughly $25\%$ for $ q = 1/3$ systems. With the exception of $q = 1$ systems as observed by H1, Eq.~\eqref{eqn:mod_criteria} tends to provide conservative estimates for SNR.

Realistic predictions for the minimum resolution required for unbiased PE for CE requires waveforms that span the entire CE frequency range. However, we are limited by the length of the NR waveforms.
While it is possible to extend the length of NR waveforms through hybridization, doing so introduces errors in addition to the truncation errors this study is isolating. 
Thus, we employ power-weighted mismatches \cite{ohme2011reliability,PhysRevD.109.044032}, instead of conventional mismatches, to calculate $\beta$ in Eq.~\eqref{eq:beta} and subsequently calculate estimates for the minimum resolution required for unbiased PE for CE. 
Power-weighted mismatches allow us to approximately account for the absent power in the 5--20 Hz range and can be computed as follows:
\begin{equation}
    \epsilon_{\text{pow}}[h_1, h_2] = \frac{\langle h | h \rangle_{(20,2048)}}{\langle h | h \rangle_{(5,2048)}}  \epsilon[h_1, h_2]\,.
    \label{eqn:power-mm}
\end{equation}
Here, $h_1, h_2$ are two NR waveforms differing only in simulation grid resolution, $\langle h | h \rangle_{(20,2048)}$ is the noise-weighted inner product integrated from 20--2048 Hz of any reasonably accurate waveform model $h$ assumed to approximate the inspiral phase, and similarly for $\langle h | h \rangle_{(5,2048)}$.
The ratio of the power in the reduced bandwidth to the full bandwidth accounts for the early inspiral absent from our analysis, which is optimistically expected to have a perfect match between $h_1$ and $h_2$ in the missing frequency band.
Here, we use \NRSur for $h$.
Since, $\epsilon_{\text{pow}} \leq \epsilon$, our $\beta$ values are lower bounds to the actual values computed with $\epsilon$.
This means that our predicted $\Delta$ values are upper bounds on the required grid spacing, hence serving as lower bounds on the required $\Delta^{-1}$ for CE.
We find that when using power-weighted mismatches, the resolution prediction for the synthetic sources in Table \ref{tab:param} for CE is approximately 95\% of what we obtain when using normal mismatches calculated in the reduced bandwidth.

\begin{figure*}
    \centering
    \includegraphics[scale = 0.6] {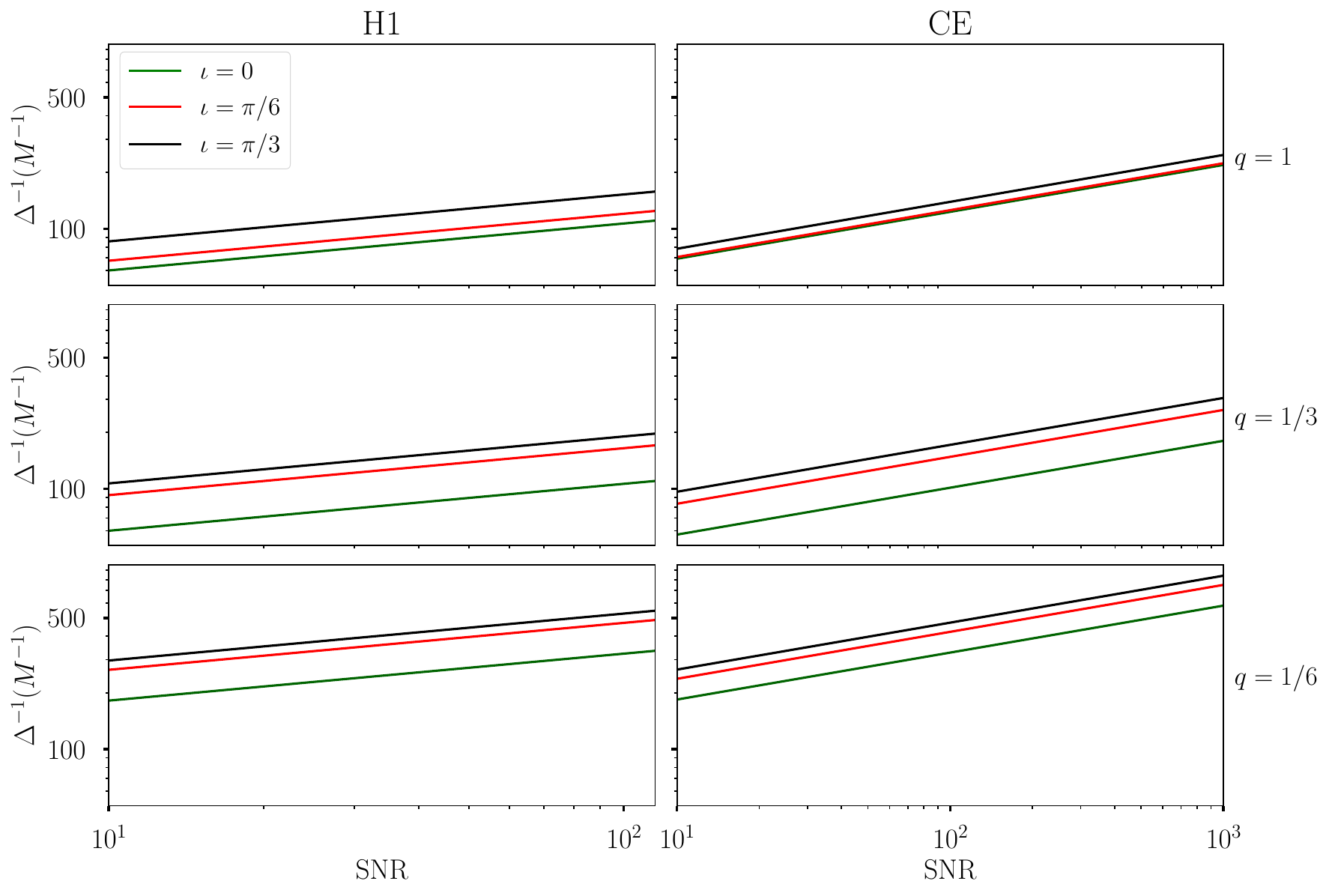}
    \caption{\textbf{Applying Eq.~\eqref{eqn:mod_criteria} to predict accuracy requirements}: The predictions for H1 are presented in the left column, while the right column displays predictions for CE. Each row corresponds to predictions for one of the three mass ratios. The colors represent the three inclinations.}
    \label{fig:prediction}
\end{figure*}

We now apply this modified criterion to $q=1$ and $q=1/3$ waveforms observed at three different inclinations, $0$, $\pi/6$ and $\pi/3$. 
We also apply this criterion to aligned $q=1/6$ {\tt MAYA} waveforms, which have $\chi_{1z} = 0.2$ and $\chi_{2z} = 0.0$. We have set $M_\text{tot} = 100 M_\odot$ for these waveforms.
Figure~\ref{fig:prediction} provides a visual representation of the results when this modified criterion is applied to these three different mass ratio systems observed at three distinct inclinations and for both H1 and CE. 
The figure illustrates that our current waveforms suffice in terms of accuracy for systems with $q = 1$ and $q=1/3$, regardless of whether we are dealing with H1 or CE. For systems with $q = 1/6$, we find that our current resolutions in the ${\tt MAYA}$ catalog will be accurate for the median SNR  of $10$ for aLIGO \cite{2021arXiv211103606T} and $20$ for CE \cite{reitze2019cosmic}. But considering that LIGO Voyager will be at least four times more sensitive than aLIGO \cite{Voyager} and CE will observe signals with SNRs above $600$, we find that current resolutions are not sufficient for such high SNR signals for inclinations greater than $\pi/3$ for H1 and for all possible inclinations for CE.

It is important to note that the accuracy prediction is dependent on the total mass since reducing $M_{\text{tot}}$ places more of the waveform within the detectors' sensitive bands, causing more truncation error to accumulate and consequently lowering the SNR at which a given resolution introduces significant parameter bias.

\section{Conclusions}
\label{sec:conclude}

In this work, we have assessed the impact of NR truncation errors on PE. We accomplished this by performing PE, across a range of SNRs, on simulated GW signals generated at different NR resolutions. We found that for SNRs where a specific resolution ensures unbiased PE, employing a higher resolution does not yield additional scientific insights. 
Additionally, we have shown that the resolution required for unbiased PE increases as the mass ratio decreases and/or the observed inclination angle increases. The accuracy requirements are also influenced by the total mass of the system; as the total mass decreases, more accurate waveforms are necessary to achieve an equivalent level of PE accuracy at a given SNR. Furthermore, the shape of the noise curve of a detector is a key factor in defining accuracy requirements. 
The accuracy demands differ between detectors such as CE and aLIGO, particularly due to the heightened sensitivity of CE at lower frequencies.

We have provided a measure for determining the SNR at which a resolution will produce significant parameter bias. By comparing the critical SNR predictions from this measure with those from full PE for aLIGO and CE, we have shown that the measure provides reasonably accurate estimates of the critical SNR, with most predictions being conservative. To make predictions for the resolution requirements for future NR codes, we applied this measure to three different mass ratio NR waveforms observed at various inclinations and for both detectors. From this application we predict, for equal and moderately unequal mass ratio, our current NR waveforms will be sufficiently accurate, even when observed at high inclinations. For mass ratios around $1/6$, our current resolutions will be accurate for the median SNRs for both detectors. However, they will introduce significant parameter bias in the PE for high SNR signals, at all inclinations for CE, and at high inclinations for LIGO Voyager. Considering that, at a given SNR, the resolution required for unbiased PE increases as the mass ratio decreases, current resolutions will produce significant parameter bias in the PE of much lower mass ratio signals even at median SNRs.

In order to attain unbiased PE of signals from binary systems with low mass ratios, we need to generate more accurate NR waveforms.
However, the generation of accurate waveforms is a time-consuming process
requiring finite-differencing codes to efficiently scale with an increased number of computational nodes.
Consequently, significant efforts are underway to improve the parallelization and scalability of NR codes \cite{foucart2022snowmass2021,lisaconsortiumwaveformworkinggroup2023waveform}. 
It is imperative to achieve these improvements to maximize the scientific outcomes of GW observations.

In the future, we plan to study the impact of other sources of NR errors, such as extraction radius and finite-differencing order, on PE. Additionally, we aim to explore the impact of truncation error on the recovery of parameters like $\mc_c$ and $\chi_\text{eff}$ over a wide range of injection parameters. We also plan to extend this analysis to the Laser Interferometer Space Antenna (LISA). Given LISA's capability to detect signals with high SNRs and across a wider range of parameters than those examined in this study, our emphasis will be directed towards studying such systems.

\begin{acknowledgements}
The work presented in this paper is possible due to NSF Grants PHY-2207780 and PHY-2114581. AZ was supported by NSF Grants PHY-2207594 and PHY-2308833.
DF was supported by NSF Grant OAC-2004879. The computing resources necessary to perform the NR simulations were provided XSEDE PHY120016 and TACC PHY20039. 
The authors are grateful for the computational resources used for the PE runs provided by the LIGO Laboratories at CIT, LHO, and LLO supported by NSF Grants PHY-0757058 and PHY-0823459. The authors are also grateful for the computational resources provided on the Nemo cluster by the Leonard E Parker Center for Gravitation, Cosmology, and Astrophysics at the University of Wisconsin-Milwaukee supported by NSF Grants PHY-1626190 and PHY-1700765. Finally, the authors acknowledge the resources used from the Sarathi cluster by the Inter-University Center for Astronomy \& Astrophysics (IUCAA), Pune, India.
This material is based upon work supported by NSF's LIGO Laboratory which is a major facility fully funded by the National Science Foundation.
The work was done by members of the Weinberg Institute and has an identifier of UT-WI-44-2023. 
\end{acknowledgements}

\appendix

\section{JS divergence}
\label{sec:JS}

\begin{figure}
    \includegraphics[width = \columnwidth]{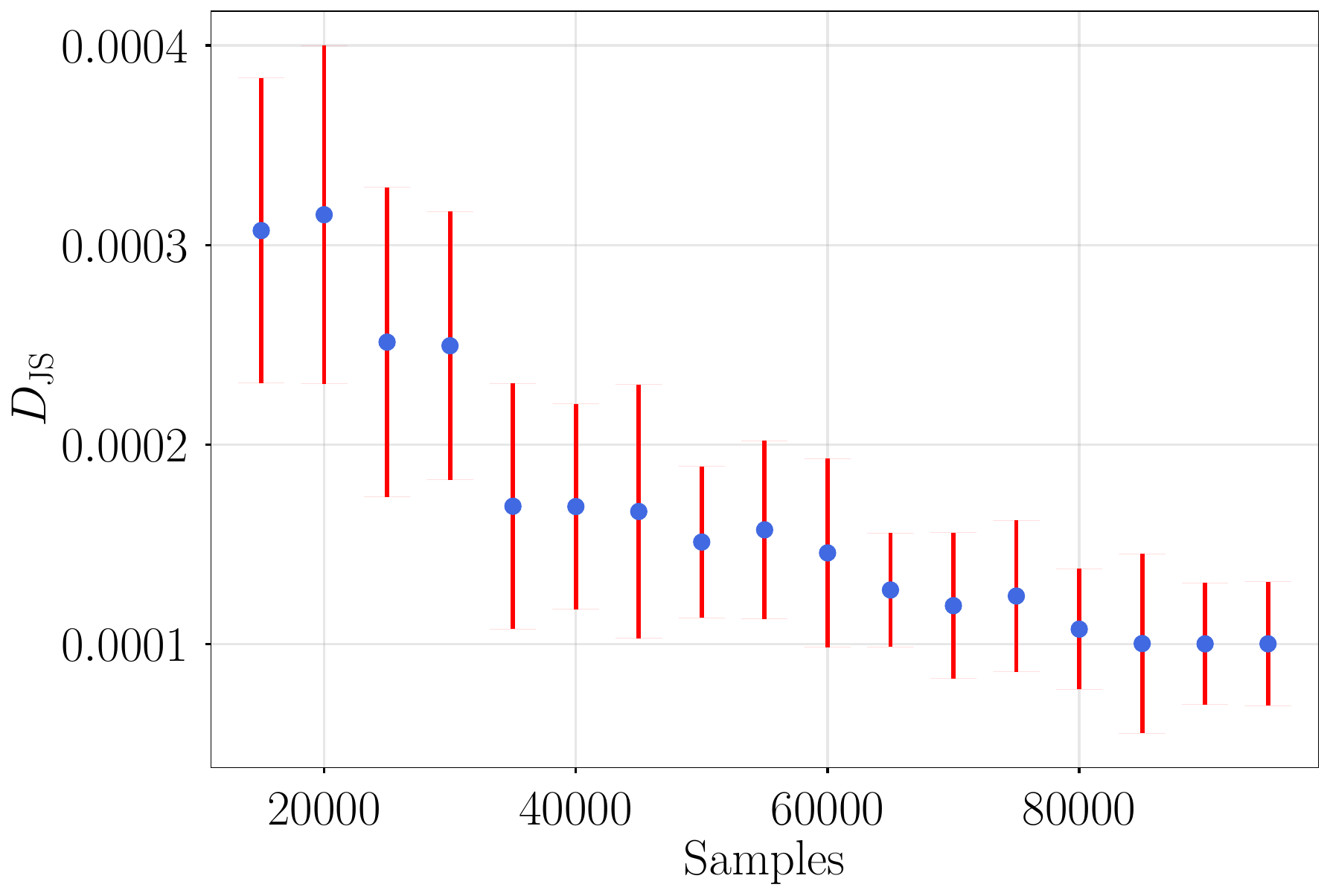}
    \caption{\textbf{Sampling errors}: JS divergence computed between KDEs built from two sets of samples from standard normal distributions
    as a function of the number of samples, illustrating the effect of sampling errors. 
    The points are medians, with error bars denoting one standard deviation. 
    These values come from 1000 JS divergence calculations at each $x$-axis point.}
    \label{fig:JSD}
\end{figure}

The JS divergence \cite{JS_test}, is a statistical tool employed to quantify the dissimilarity between two posterior distributions.  
It is convenient to use when assessing the agreement between these distributions as it is symmetric in nature, meaning $D_\text{JS}(a|b) = D_\text{JS}(b|a)$ and its output is bounded between $0$ and $1$ bit, where $0$ bit means the two distributions are identical and $1$ bit means maximal divergence \cite{10.1093/mnras/staa2850}. 

Given two discrete probability distributions $a(x)$ and $b(x)$, the JS divergence  between them is defined as:
\begin{align}
    D_\text{JS}(a|b) = \frac{1}{2} \int \biggl[ 
     a(x) & \log_2\biggl( \frac{a(x)}{m(x)}\biggr)
    \notag\\  
    &
     + b(x)\log_2\biggl( \frac{b(x)}{m(x)}\biggr) \biggr]dx
\label{eq:JS_test}
\end{align}
where  $m(x)=[a(x)+b(x)]/2$. To evaluate the convergence of the {\tt RIFT} algorithm, we generate kernel density estimators (KDEs) for the marginal probability density functions (PDFs) for each intrinsic parameter. These estimators are constructed using the samples obtained from two consecutive iterations and are subsequently utilized as $a$ and $b$ in Eq.~\ref{eq:JS_test} to calculate the JS divergence between them. Our chosen convergence criterion entails achieving a JS divergence of less than $10^{-3}$ between two successive iterations. This threshold is determined while taking into account the anticipated JS divergence resulting solely from sampling errors. Consider the following experiment: we draw two sets of independent samples from standard normal distributions. 
We then carry out our JS procedure with these samples, forming KDEs for $a$ and $b$ from the two sets and computing $D_{\rm JS}(a|b)$.
In Fig.~\ref{fig:JSD} we illustrate the median and standard deviation of $D_{\rm JS}$ from 1000 iterations of this test as we vary the number of samples.
JS divergence values of $\sim 10^{-4}$ are thus expected due to sampling errors, even for identical distributions.
In our study, we typically use 85,000 samples from {\tt RIFT} to test convergence, to ensure that there is no impact from sampling variance.

\section{Additional results}
\label{sec:FurtherResults}

\subsection{$\boldsymbol{q=1}, \boldsymbol{\iota = \pi/6}$}
\label{assec:q1_incl30}

\begin{figure}
    \centering
    \includegraphics[width = \columnwidth]{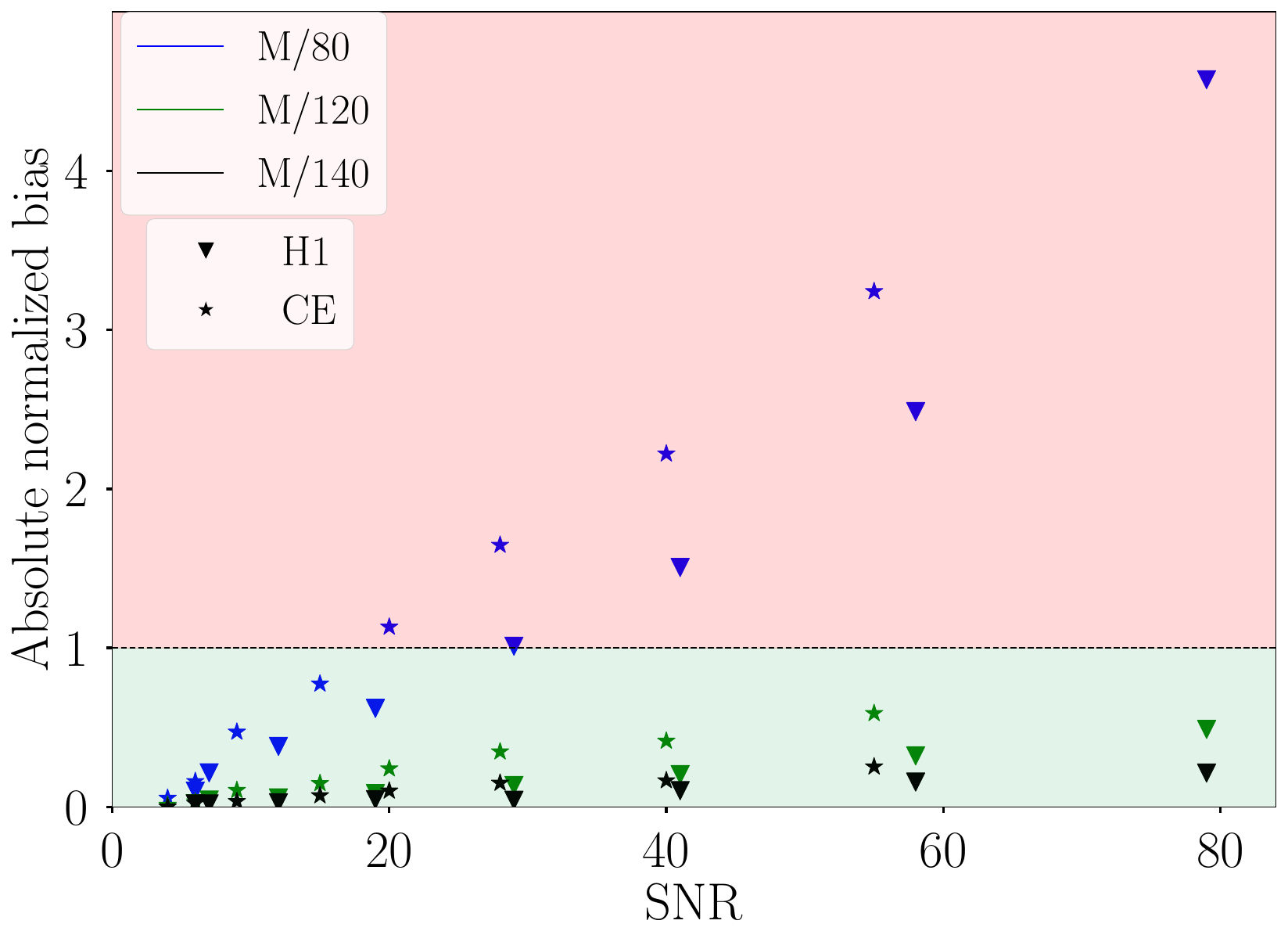}
   \caption{\textbf{Absolute normalized bias in the marginalized $\mathcal{M}_c$ posterior distributions of $\boldsymbol{q=1, \iota = \pi/6}$ injections as a function of SNR}: The dashed horizontal line marks an absolute normalized bias of unity. The bias values were calculated with respect to $M/200$.}
    \label{fig:bias_q1_incl30}
\end{figure}

\begin{figure*}
    \includegraphics[width=\columnwidth]{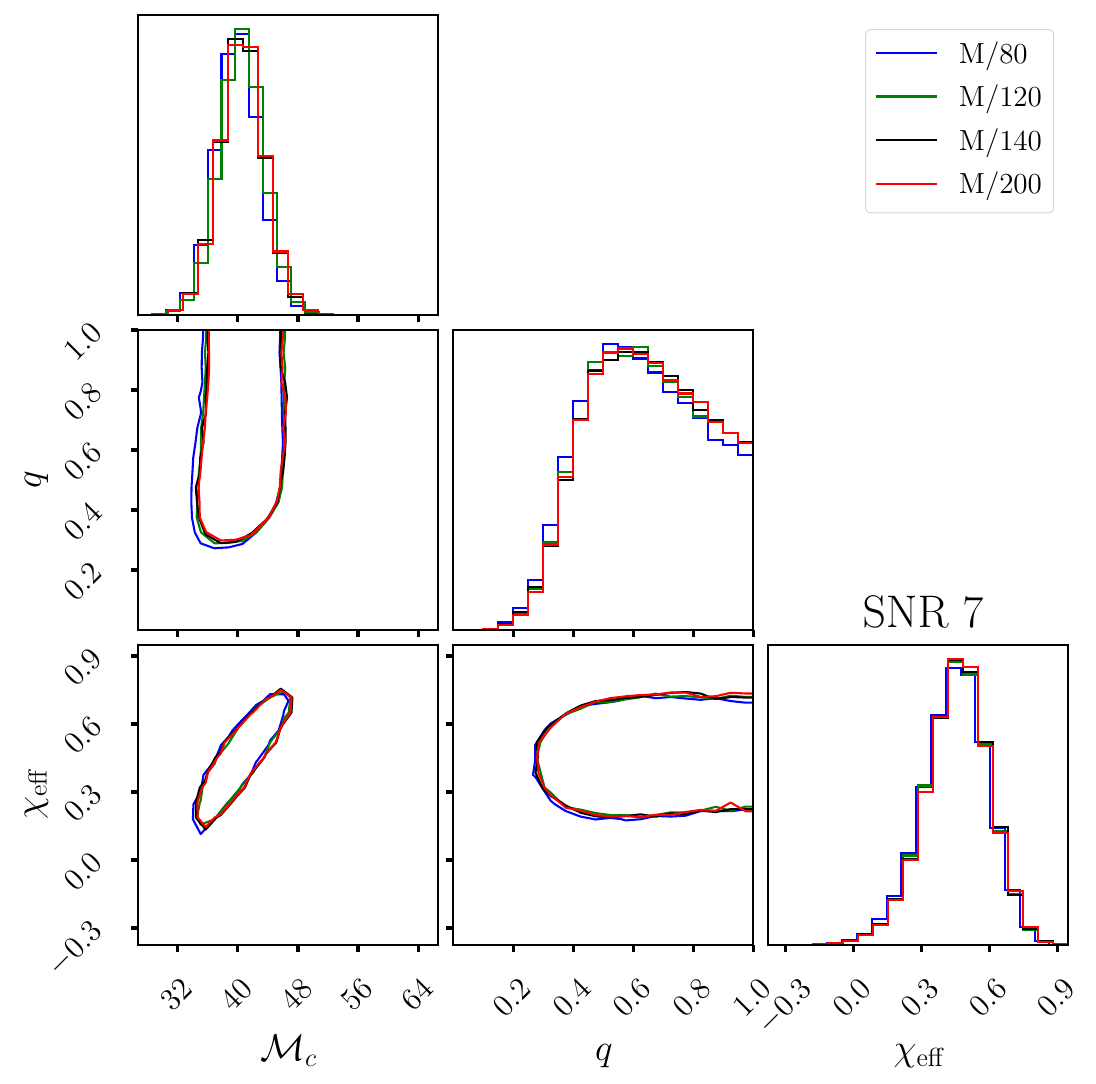}
    \includegraphics[width=\columnwidth]{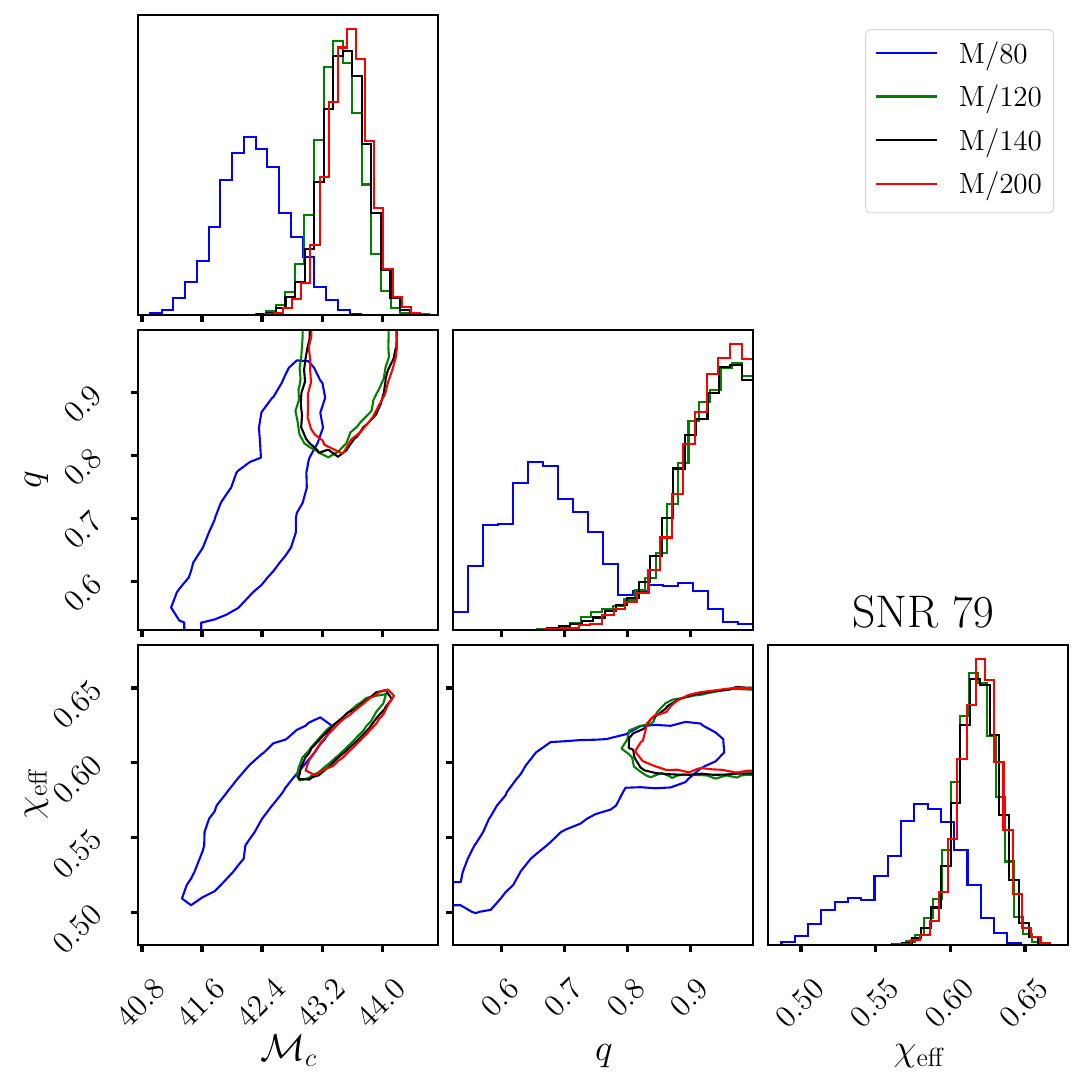}
    \includegraphics[scale = 0.5] {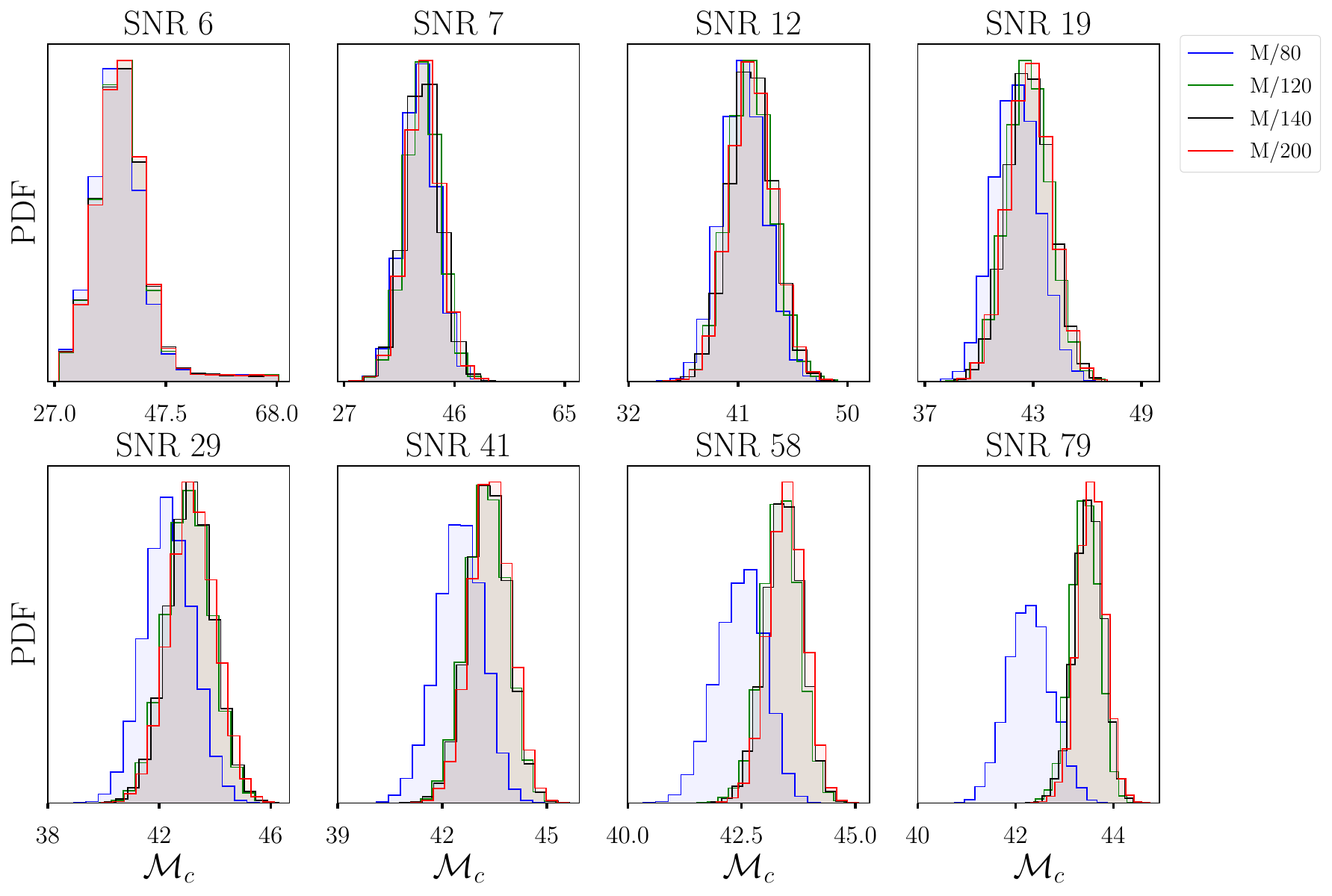}
    \caption{\textbf{PE results for $\boldsymbol{q=1, \iota = \pi/6}$ injections (H1)}: \textit{Top left}: One- and two-dimensional marginal posterior distributions for $\mathcal{M}_c, q,$ and $\chi_{\text{eff}}$. Diagonal panels show the one-dimensional marginal posterior distribution, while contours in the off-diagonal panels show the 90\% credible intervals for the two-dimensional marginal posterior distribution. Different colored curves correspond to different resolutions. Injections had an SNR of $7$ and the minimum resolution for indistinguishability at that SNR is predicted to be $(M/75)^{-1}$. \textit{Top right:} Corner plot produced after performing PE at an SNR of $79$, where the minimum resolution for indistinguishability is predicted to be $(M/135)^{-1}$.
    \textit{Bottom}: One-dimensional marginalized posterior distributions for $\mc_c$ are presented here. PE was conducted at a sequence of SNRs, with all parameters held constant except for $D_L$. Each panel illustrates the outcomes for a specific SNR, and distinct colored curves represent different resolutions. With increasing SNR, the $M/80$ posterior gradually separates from the others.}
    \label{fig:corner_H1_q1_incl30}
\end{figure*}

\begin{figure*}
    \includegraphics[width=\columnwidth]{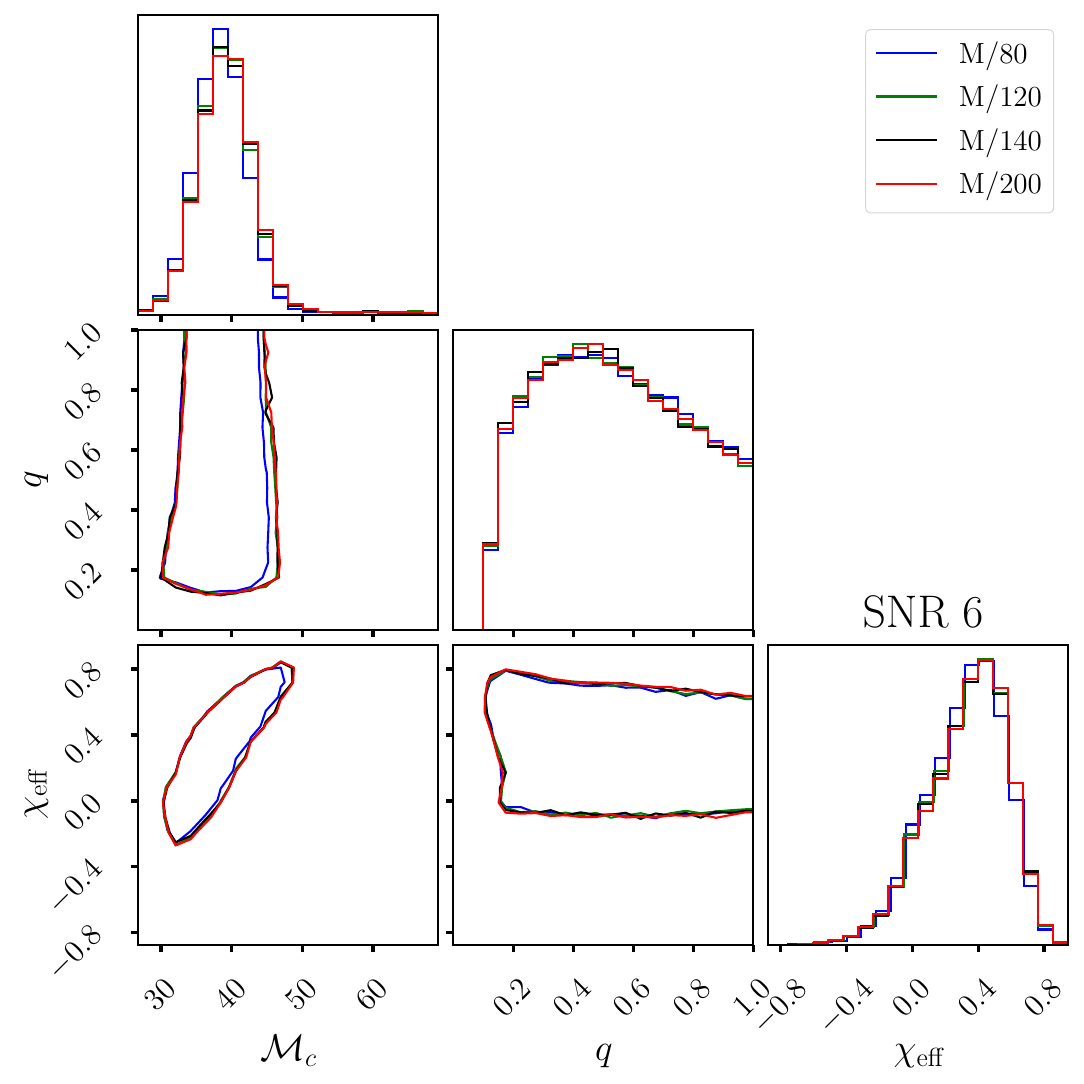}
    \includegraphics[width=\columnwidth]{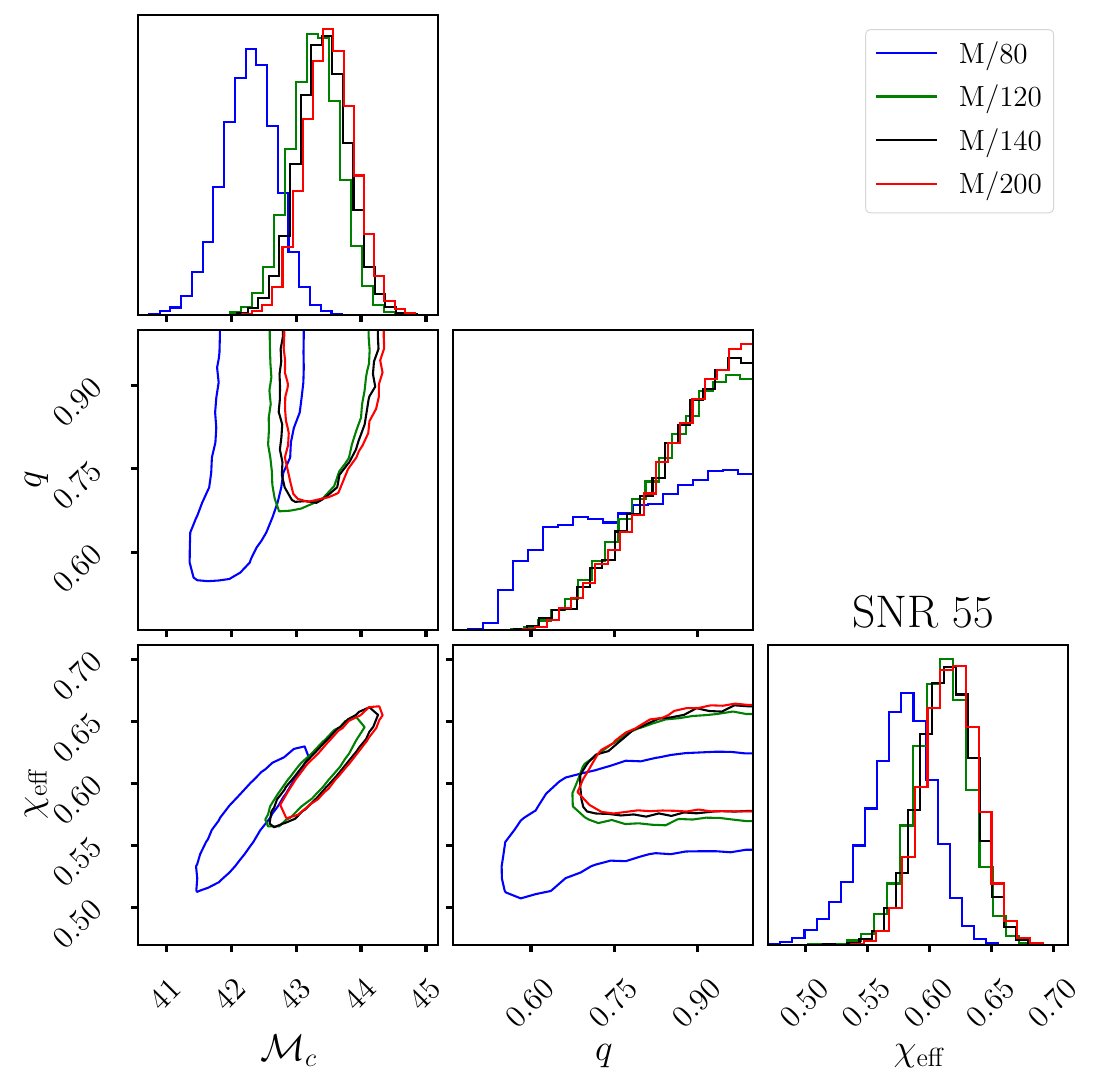}
    \includegraphics[scale = 0.5]{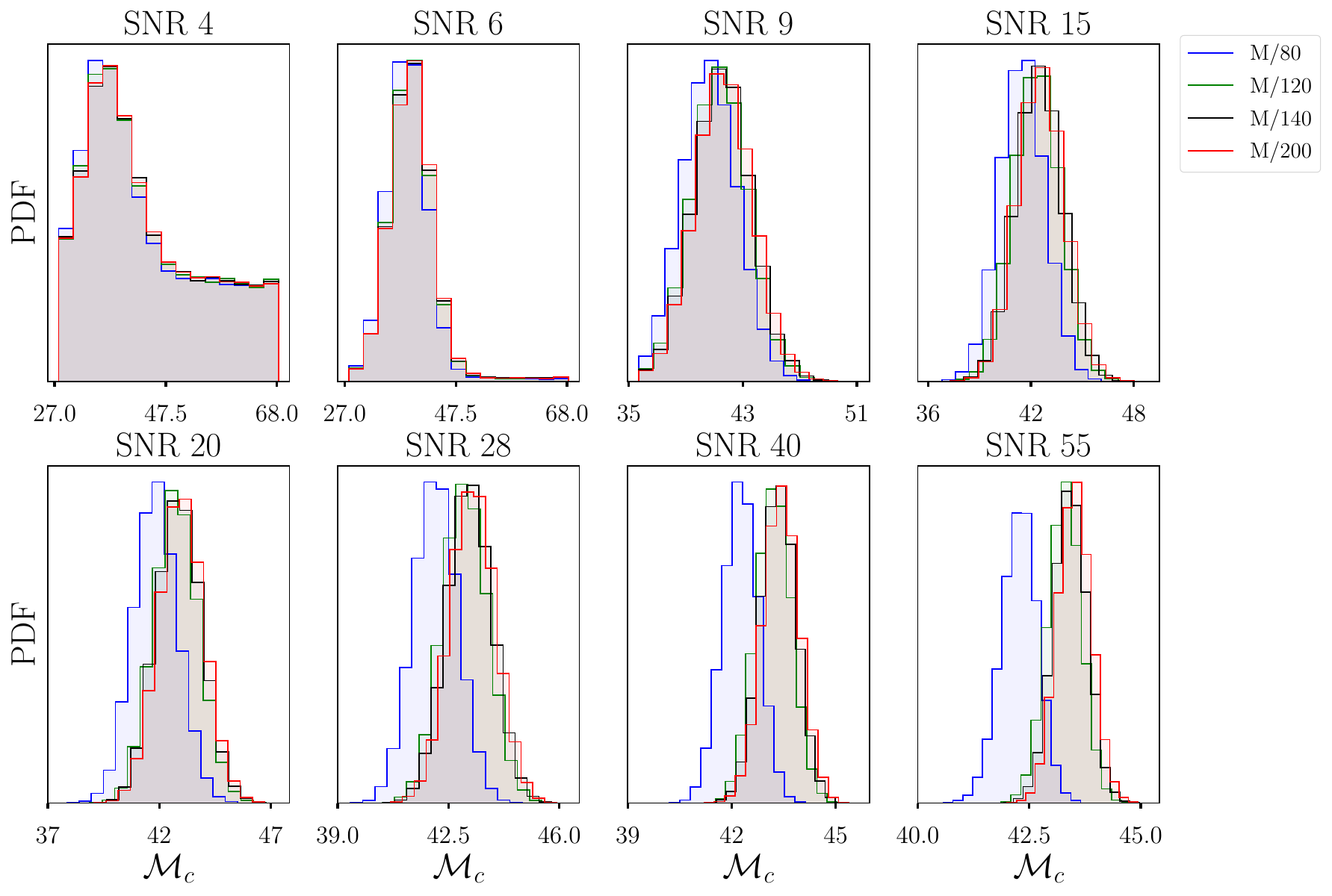}
    \caption{\textbf{PE results for $\boldsymbol{q=1, \iota = \pi/6}$ injections (CE)}: 
    \textit{Top left}: One- and two-dimensional marginal posterior distributions for $\mathcal{M}_c, q,$ and $\chi_{\text{eff}}$. Diagonal panels show the one-dimensional marginal posterior distribution, while contours in the off-diagonal panels show the 90\% credible intervals for the two-dimensional marginal posterior distribution. 
    Different colored curves correspond to different resolutions. Injections had an SNR of $6$ and the minimum resolution for indistinguishability at that SNR is predicted to be $(M/75)^{-1}$. 
    \textit{Top right:} Corner plot produced after performing PE at an SNR of $55$, where the minimum resolution for indistinguishability is predicted to be $M(/135)^{-1}$.
    \textit{Bottom}: One-dimensional marginalized posterior distributions for $\mc_c$ are presented here. PE was conducted at a sequence of SNRs, with all parameters held constant except for $D_L$. Each panel illustrates the outcomes for a specific SNR, and distinct colored curves represent different resolutions. With increasing SNR, the $M/80$ posterior gradually separates from the others.}
    \label{fig:corner_CE_q1_incl30}
\end{figure*}

Fig.~\ref{fig:bias_q1_incl30} provides a comprehensive summary of our findings for this system, mirroring the observations made in the face-on case. Additionally, the PE results  are provided in Fig.~\ref{fig:corner_H1_q1_incl30} (Fig.~\ref{fig:corner_CE_q1_incl30}) for H1 (CE). 
The top panel shows the 1-D and 2-D histogram plots for SNR $7$ ($6$) and SNR $79$ ($55$), where $M/75$ and $M/135$ are the predicted critical grid spacings respectively. 
The bottom panel shows the detector-frame $\mathcal{M}_c$ posterior distribution for all eight SNRs and Table \ref{tab:q1_incl30_bias} provides the normalized bias values as a function of SNR.

\begin{table*}
\centering
\addtolength{\tabcolsep}{20pt} 
     \begin{tabular}{c c c c c}
     \hline
     SNR & $\Delta_\text{critical}$ & $M/80$ &$M/120$ & $M/140$\\ [0.5ex] 
     \hline\hline
     6 & $M/70$ & -0.10 & -0.02 & -0.02  \\ 
     7 & $M/75$ & -0.21 & -0.04 & -0.02 \\
     12 & $M/85$ & -0.38 & -0.06 & -0.03 \\
     19 & $M/95$ & -0.62 & -0.08 & -0.05 \\
     29 & $M/105$  & -1.01 & -0.13 & -0.04 \\
     41 & $M/115$  & -1.51 & -0.20 & -0.10 \\
     58 & $M/125$ & -2.49 & -0.32 & -0.16 \\
     79 & $M/135$ & -4.57 & -0.49 & -0.21 \\[1ex] 
     \hline \\

    \hline
     SNR & $\Delta_\text{critical}$ & $M/80$ &$M/120$ & $M/140$\\ [0.5ex] 
     \hline\hline
     4 & $M/70$ & -0.06 & -0.01 & -0.00  \\ 
     6 & $M/75$ & -0.16 & -0.03 & -0.02 \\
     9 & $M/85$ & -0.47 & -0.11 & -0.04 \\
     15 & $M/95$ & -0.77 & -0.15 & -0.07 \\
     20 & $M/105$  & -1.13 & -0.24 & -0.10 \\
     28 & $M/115$  & -1.65 & -0.35 & -0.15 \\
     40 & $M/125$ & -2.22 & -0.41 & -0.17 \\
     55 & $M/135$ & -3.24 & -0.59 & -0.25 \\[1ex] 
     \hline \\
     \end{tabular}
     \addtolength{\tabcolsep}{-20pt}
     \caption{\textbf{Normalized bias in the marginalized $\mc_c$ posterior distributions of $\boldsymbol{q=1, \iota = \pi/6}$ injections}: Bias observed in the lower resolution posteriors, calculated with respect to $M/200$, for H1 (top) and CE (bottom).}
    \label{tab:q1_incl30_bias}
\end{table*}

\subsection{$\boldsymbol{q=1/3}$, $\boldsymbol{\iota = 0}$}
\label{assec:q3}

\begin{figure*}
    \includegraphics[width=\columnwidth]{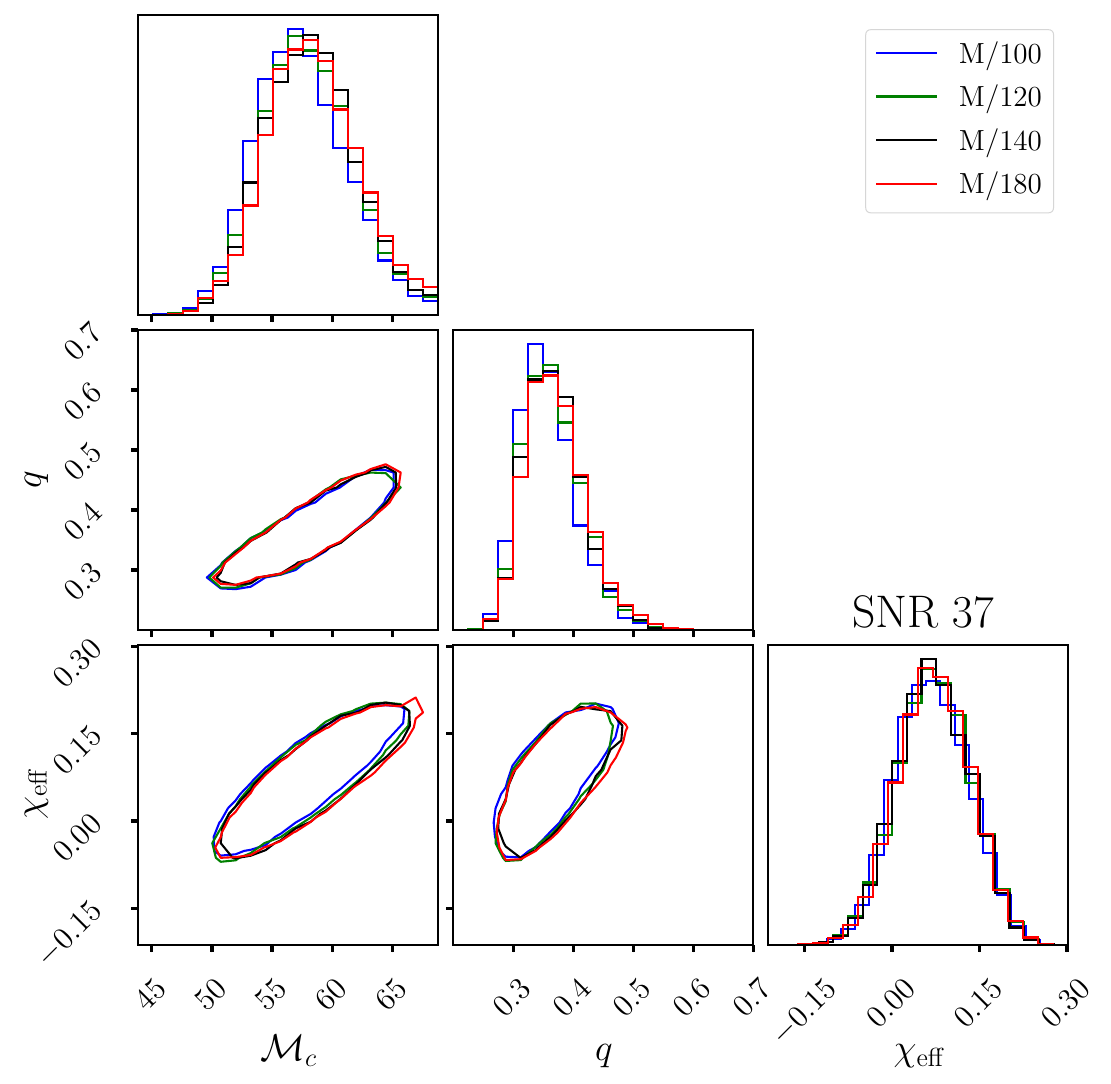}
    \includegraphics[width=\columnwidth]{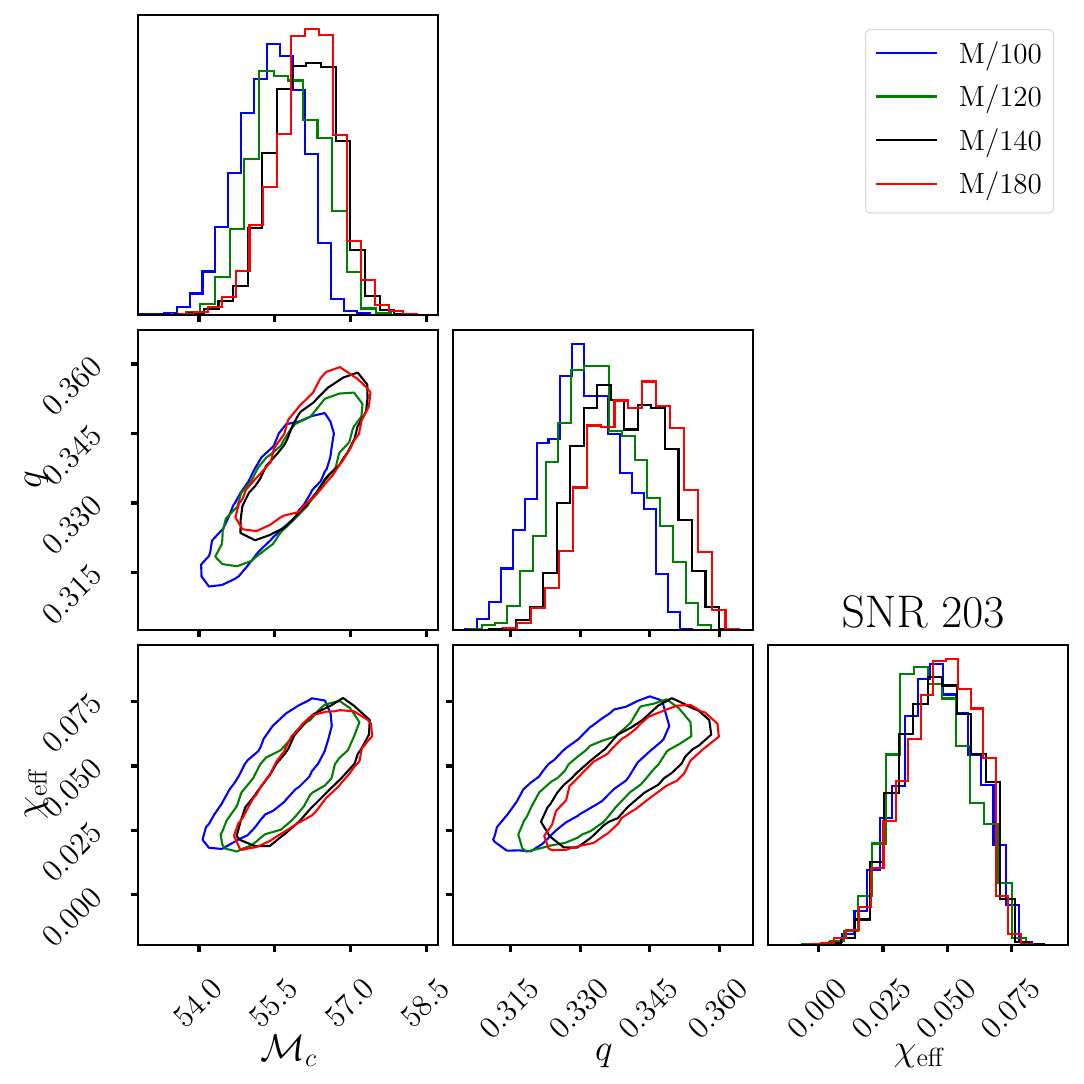}
    \includegraphics[scale = 0.5] {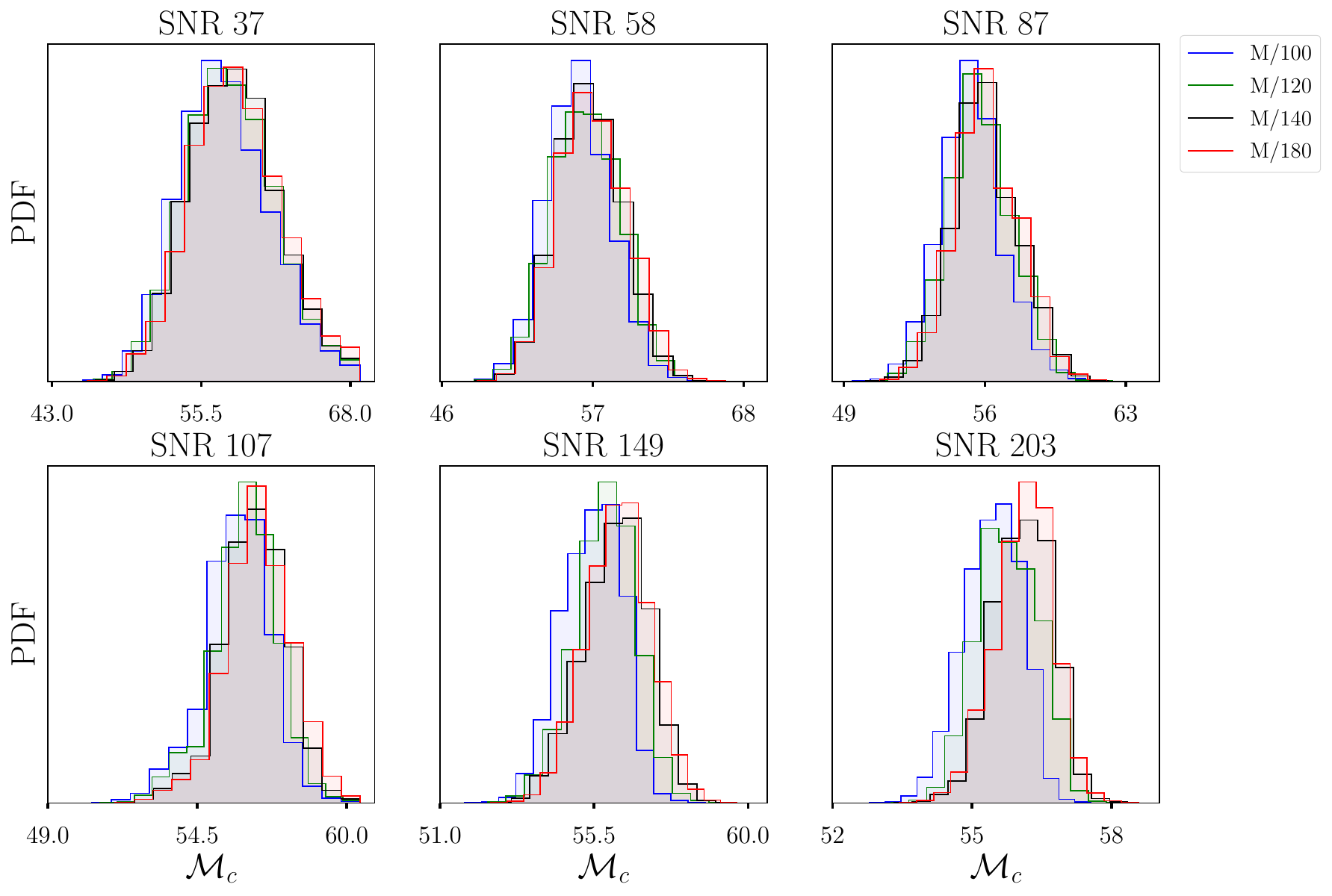}
    \caption{\textbf{PE results for $\boldsymbol{q=1/3, \iota = 0}$ injections (H1)}: \textit{Top left}: One- and two-dimensional marginal posterior distributions for $\mathcal{M}_c, q,$ and $\chi_{\text{eff}}$. Diagonal panels show the one-dimensional marginal posterior distribution, while contours in the off-diagonal panels show the 90\% credible intervals for the two-dimensional marginal posterior distribution. Different colored curves correspond to different resolutions. Injections had an SNR of $37$ and the minimum resolution for indistinguishability at that SNR is predicted to be $(M/85)^{-1}$. \textit{Top right:} Corner plot produced after performing PE at an SNR of $203$, where the minimum resolution for indistinguishability is predicted to be $(M/135)^{-1}$.
    \textit{Bottom}: One-dimensional marginalized posterior distributions for $\mc_c$ are presented here. PE was conducted at a sequence of SNRs, with all parameters held constant except for $D_L$. Each panel illustrates the outcomes for a specific SNR, and distinct colored curves represent different resolutions}
    \label{fig:corner_H1_q3}
\end{figure*}

\begin{figure*}
    \includegraphics[width=\columnwidth]{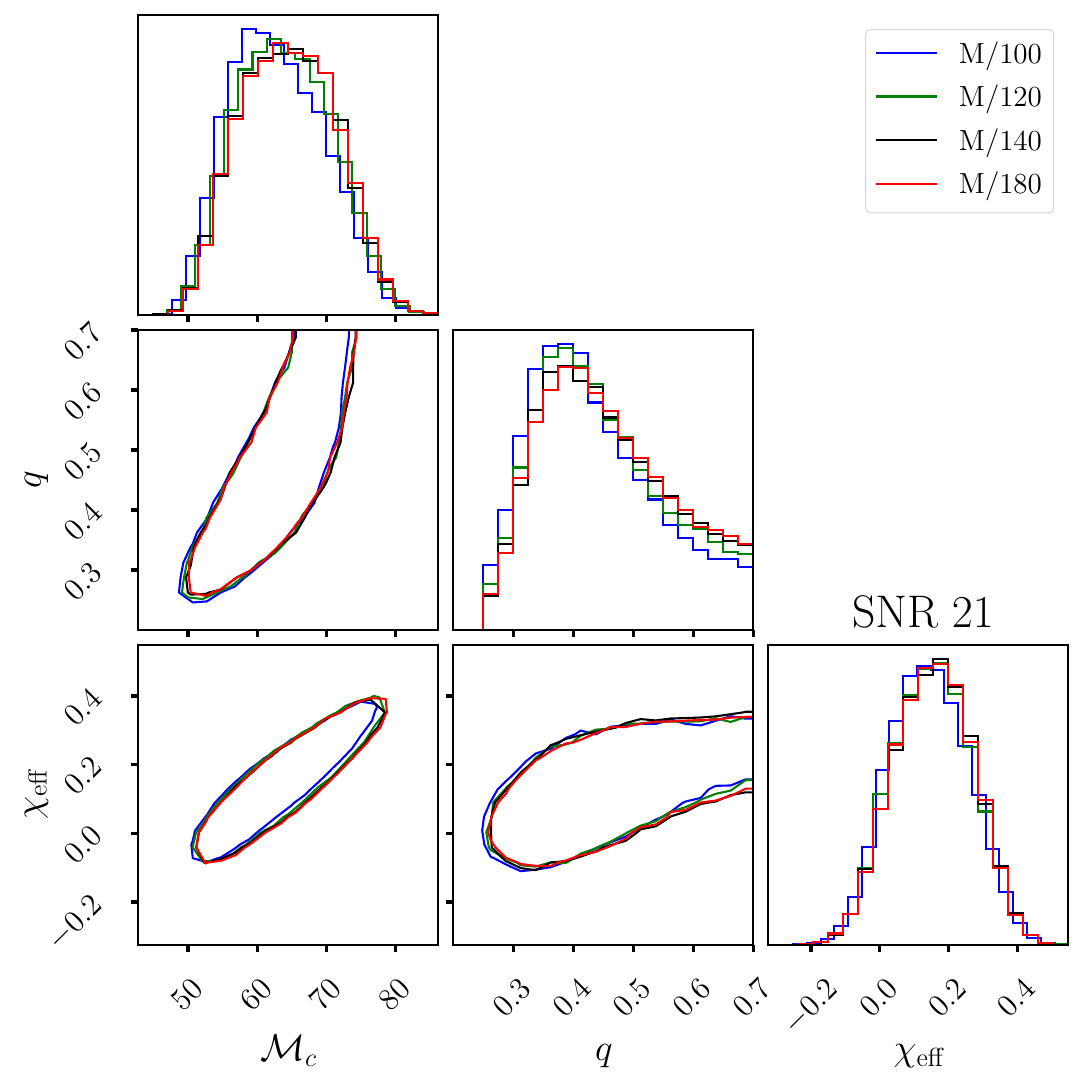}
    \includegraphics[width=\columnwidth]{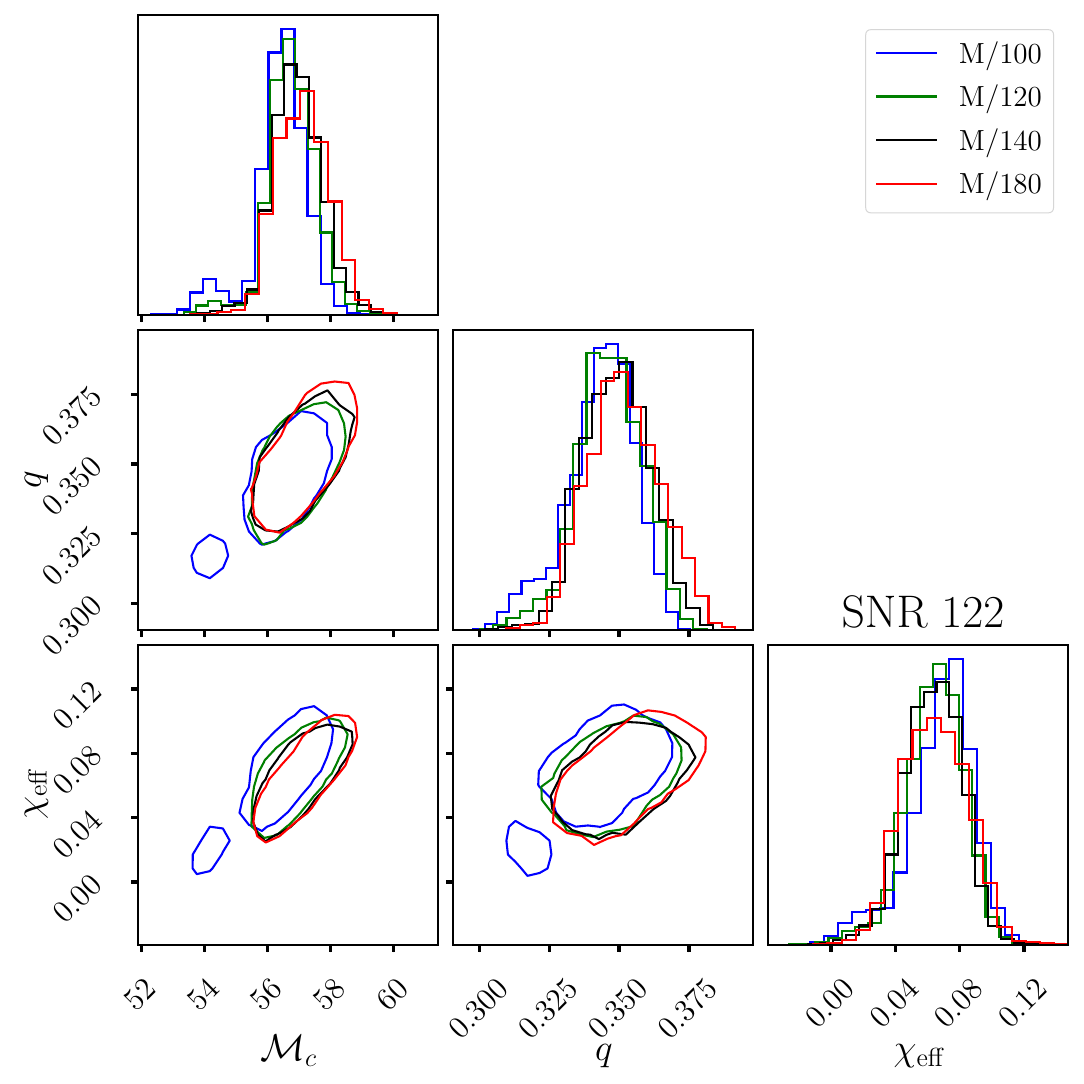}
    \includegraphics[scale = 0.5]{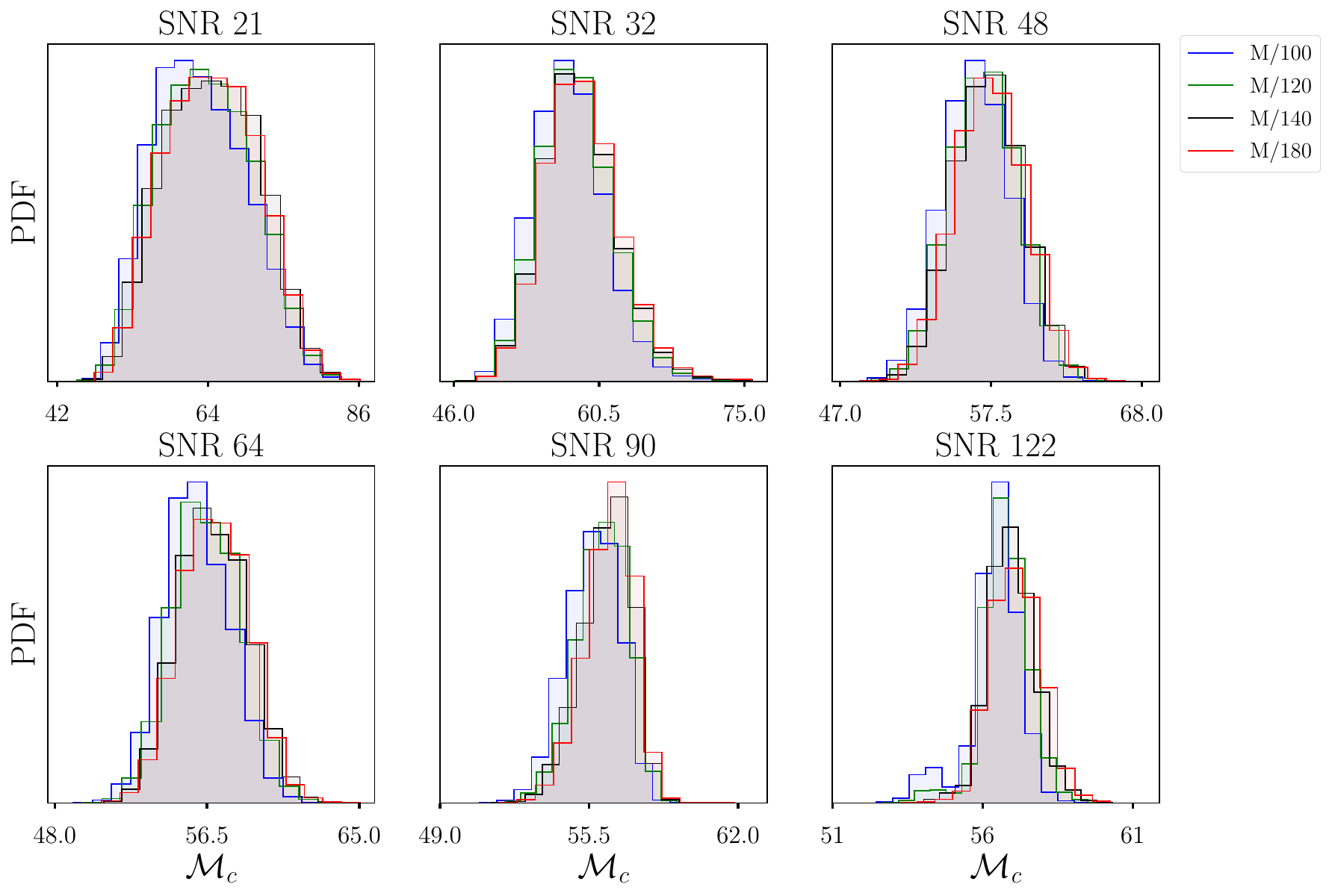}
    \caption{\textbf{PE results for $\boldsymbol{q=1/3, \iota = 0}$ injections (CE)}: \textit{Top left}: One- and two-dimensional marginal posterior distributions for $\mathcal{M}_c, q,$ and $\chi_{\text{eff}}$. Diagonal panels show the one-dimensional marginal posterior distribution, while contours in the off-diagonal panels show the 90\% credible intervals for the two-dimensional marginal posterior distribution. Different colored curves correspond to different resolutions. Injections had an SNR of $21$ and the minimum resolution for indistinguishability at that SNR is predicted to be $(M/85)^{-1}$. \textit{Top right:} Corner plot produced after performing PE at an SNR of $122$, where the minimum resolution for indistinguishability is predicted to be $(M/135)^{-1}$.
    \textit{Bottom}: One-dimensional marginalized posterior distributions for $\mc_c$ are presented here. PE was conducted at a sequence of SNRs, with all parameters held constant except for $D_L$. Each panel illustrates the outcomes for a specific SNR, and distinct colored curves represent different resolutions.}
    \label{fig:corner_CE_q3}
\end{figure*}

PE results are shown in Fig.~\ref{fig:corner_H1_q3} for H1 and Fig.~\ref{fig:corner_CE_q3} for CE. 
In both figures, the top panels display the 1-D and 2-D histogram
plots at SNR $37$ ($21$) and SNR $203$ ($122$) for H1 (CE), where $M/85$ and $M/135$ are the predicted critical grid spacings respectively.
While the $M/100$ posterior shows deviation from the $M/180$ posterior, the deviation is not as pronounced as was for $q=1$ injections. 
This is due to the resolutions being close to each other, minimizing the disparity.
The bottom panel in both figures shows the marginalized $\mc_c$ posteriors for the sequence of SNRs. In this panel, a similar observation is made, the lowest resolution posterior distribution deviates from the highest resolution posterior distribution, although not as prominently as observed for $q = 1$ injections. Table \ref{tab:q3_bias} provides the normalized bias values as a function of SNR for both detectors.

\begin{table*}
\centering

\addtolength{\tabcolsep}{20pt} 
     \begin{tabular}{c c c c c}
     \hline
     SNR & $\Delta_\text{critical}$  & $M/100$ &$M/120$ & $M/140$\\ [0.5ex] 
     \hline\hline
     37 & $M/85$ & -0.28 & -0.12 & -0.03  \\ 
     58 & $M/95$ & -0.39 & -0.16 & -0.06 \\
     87 & $M/105$ & -0.41 & -0.15 & -0.01 \\
     107 & $M/115$ & -0.49 & -0.35 & -0.08 \\
     149 & $M/125$  & -0.78 & -0.38 & -0.05 \\
     203 & $M/135$  & -1.11 & -0.65 & -0.06 \\[1ex] 
     \hline \\

    \hline
     SNR & $\Delta_\text{critical}$ & $M/100$ &$M/120$ & $M/140$\\ [0.5ex] 
     \hline\hline
     21 & $M/85$ & -0.31 & -0.12 & -0.01  \\ 
     32 & $M/95$ & -0.34 & -0.15 & -0.06 \\
     48 & $M/105$ & -0.42 & -0.17 & -0.05 \\
     64 & $M/115$ & -0.58 & -0.26 & -0.09 \\
     90 & $M/125$  & -0.69 & -0.31 & -0.09 \\
     122 & $M/135$  & -0.76 & -0.46 & -0.24\\[1ex] 
     \hline
     \end{tabular}
     \addtolength{\tabcolsep}{-20pt}
     \caption{\textbf{Normalized bias in the marginalized $\mc_c$ posterior distributions of $\boldsymbol{q=1/3, \iota = 0}$ injections}: Bias observed in the lower resolution posteriors, calculated with respect to $M/180$, for H1 (top) and CE (bottom).}
    \label{tab:q3_bias}
\end{table*}

\pagebreak

\bibliography{references}

\end{document}